\documentclass[12pt, reqno]{amsart}
\usepackage{graphicx,amsmath,amssymb,epsfig,natbib,setspace}
\usepackage{amsfonts}
\theoremstyle{plain}
\newtheorem{theorem}{Theorem}

\newtheorem{corollary}{Corollary}

\newtheorem{lemma}{Lemma}

\newtheorem{proposition}{Proposition}
\theoremstyle{definition}
\newtheorem{remark}{Remark}

\numberwithin{equation}{section}
\newcommand{\Gn}{\mathbb{G}_n}
\newcommand{\Gp}{\mathbb{G}}

\newcommand{\Pp}{\mathrm{P}}
\newcommand{\Ep}{{\mathrm{E}}}

\newcommand{\mQ}{\mathcal{Q}}

\newcommand{\ssum}{{\textstyle \sum}}

\begin{document}
\onehalfspacing

\title[ ]{Inference on Sets in Finance}
\author[ ]{Victor
Chernozhukov$^\dag$  \ \ Emre Kocatulum$^\S$ \
\ Konrad Menzel$^\ddag$   } \noindent \date{March 2008 - this version: \today.  This paper
was first presented at the March 2008 CEMMAP conference on Inference in Partially
Identified Models with Applications. We thank Enrique Sentana and Anna Mikusheva for pointing
the relevance of structured projection approach as a benchmark for comparison, and Isaiah Andrews,
Dan Greenwald, and Ye Luo for detailed comments on various drafts of the paper. We also
thank Raj Chetty, Patrick Gagliardini, Raffaella Giacomini, Jerry Hausman, Olivier Scaillet, Nour Meddahi, Enno Mammen and participants
of seminars at MIT, CEMMAP, 2009 Tolouse Conference on Financial Econometrics, and 2009 European Econometric Society Meetings in Milan for useful discussion and comments. }
 \maketitle

\begin{abstract} In this paper we consider the problem of inference on  a class of sets describing a collection of admissible models as solutions to a single smooth inequality. Classical and recent examples include, among others, the Hansen-Jagannathan (HJ) sets of admissible stochastic discount factors,  Markowitz-Fama (MF) sets of mean-variances for asset portfolio returns,  and the set of structural elasticities in Chetty (2012)'s analysis
of demand with optimization frictions.   We show that the econometric structure of the problem allows us to construct convenient and powerful confidence regions based upon the weighted likelihood ratio and weighted Wald (directed weighted Hausdorff) statistics.   The statistics we formulate differ (in part) from existing statistics in that they enforce either exact or first order equivariance to transformations of parameters, making them especially appealing in the target applications.  Moreover, the resulting inference procedures are also more powerful than the structured
 projection methods, which rely upon building confidence sets for the frontier-determining sufficient parameters (e.g. frontier-spanning portfolios), and then projecting them to obtain confidence sets for HJ sets or MF sets.  Lastly, the framework we put forward is also useful for analyzing intersection bounds, namely sets defined as solutions to multiple smooth inequalities, since multiple inequalities can be conservatively approximated by a single smooth inequality. We present two empirical examples that show how the new econometric methods are able to generate sharp economic conclusions. \\[8pt]

\noindent\textbf{Keywords}: Hansen-Jagannathan bound, Markowitz-Fama bounds, Chetty bounds, Mean-Variance sets, Optimization Frictions, Inference, Confidence Set\\

\textbf{JEL classification}: C10, C50
\end{abstract}



\section{Introduction}
In this paper we consider the problem of inference on  a class of sets describing a collection of admissible models as solutions to a single smooth inequality:
\begin{equation}\label{general}
\Theta_0 = \{ \theta \in \Theta: m(\theta) \leq 0\},
\end{equation}
where $\Theta \subset \Bbb{R}^k$ is a compact parameter space, and $m: \Bbb{R}^k \mapsto \Bbb{R}$ is the inequality-generating smooth function,
which is estimable from the data.  This structure arises in a number of important examples in financial economics and public finance, as we explain below via a sequence of examples.  We show that this structure leads to highly-tractable and powerful inference procedures, and demonstrate their usefulness in substantive empirical examples. Furthermore, the structure could be used to conservatively approximate more complicated problems, where the sets of interest are given as intersections of solutions to multiple smooth inequalities. (Indeed, in the latter case, we  can conservatively approximate the multiple inequalities by a single smooth inequality. The benefits from doing so is the highly tractable, ``regular" inference.)

\subsubsection*{Example 1: Mean-Variance Set for the Stochastic Discount Factor}
In order to describe the problem, we recall \cite{cochrane}'s assertion that the science of asset pricing could be effectively summarized using the following two equations:
\begin{eqnarray*}
P_t & = & \Ep_t[ M_{t+1} X_{t+1}], \\ M_{t+1} & = &
f(Z_{t+1}, parameters),
\end{eqnarray*}
where $P_t$ is an asset price, $X_{t+1}$ is the asset payoff, $M_{t+1}$ is the stochastic discount factor (SDF) or pricing kernel (PK), which is a function $f$ of some data $Z_{t+1}$ and parameters, and $\Ep_t$ is the conditional expectation given information at time $t$. The set of SDFs $M_t$ that can price existing assets generally form a proper set, i.e. a set that is not a singleton. SDFs are not unique as long as the existing payoffs to assets do not span the entire universe of possible random payoffs. Dynamic asset pricing models provide families of potential SDFs, for example, the standard consumption model predicts that an appropriate SDF can be stated in terms of the intertemporal marginal rate of substitution:
$$
M_t =\beta\frac{u'(C_{t+1})}{u'(C_t)},
$$
where $u$ denotes a utility function parameterized by some parameters, $C_t$ denotes consumption at time $t$, and $\beta$ denotes the subjective discount factor. Note that the investor's optimal consumption plan, and therefore the marginal rate of substitution, generally depend on a set of additional state variables $Z_t$, including lifetime wealth and non-capital income. Hence, when markets are incomplete, asset prices alone do not in general pin down a unique stochastic discount factor across different aggregate states of the economy or across investors with different values of $Z_t$.

An important empirical problem in the context of the CAPM is to check which families of SDFs price the assets correctly and which do not. This reasoning forms the basis for many approaches to estimation and tests for particular specifications of an asset pricing model, most prominently \cite{HS1982}'s seminal analysis of the consumption-based CAPM, see also \cite{Lud2012} for a survey of recent developments on the subject. One leading approach for performing the check is to see whether the mean and standard deviation of SDFs $(\mu, \sigma)$ of the stochastic discount factor $M_t$ are admissible. Let $K$ be a compact convex body in $\mathbb{R}^2$. The set of admissible means and standard deviations in $K$
$$
\Theta_0 := \{ \text{ admissible pairs } (\mu,
\sigma^2) \in \Bbb{R}^2 \cap K \},
$$ is introduced by \cite{HJ1991} and known as the Hansen-Jagannathan (HJ) set.
The boundary of that set $\partial \Theta_0$ is known as the HJ bound.
In order to describe the admissible pairs in a canonical setting, let $v$ and $\Sigma$ denote the vector of mean returns and
covariance matrix to assets $1,..., N$, which are assumed not to vary with information sets at each period $t$. In the following, we consider the ``minimal sufficient parameters":
\begin{equation} \label{define ABC} S_{vv}=v^{\prime }\Sigma ^{-1}v, \;S_{v1}=v^{\prime }\Sigma ^{-1}1_{N},\; S_{11}=1_{N}^{\prime }\Sigma ^{-1}1_{N},
 \end{equation} where $1_{N}$ is a column vector of ones. Then the minimum variance $\sigma_{HJ}^2(\mu)$
achievable by a SDF given mean $\mu $ of the SDF is equal to
$$
\sigma_{HJ}^{2}\left( \mu \right) =\left( 1-\mu
v\right) ^{\prime }\Sigma ^{-1}\left( 1-\mu
v\right) =S_{vv}\mu ^{2}-2S_{v1}\mu +S_{11}.
$$
Therefore, the HJ set is equal to
$$
\Theta_{HJ} = \{ \underbrace{(\mu, \sigma)}_{\theta}
\in \underbrace{(\Bbb{R}\times\Bbb{R}_+) \cap K}_{\Theta}:
\underbrace{\sigma_{HJ}(\mu) - \sigma}_{m(\theta)}
\leq 0 \}.
$$
That is,
$$
\Theta_0 = \{ \theta \in \Theta: m_{HJ}(\theta) \leq
0\}.
$$
Note that the inequality-generating function $m_{HJ}(\theta)$ depends on the unknown means $\mu$ and covariance of returns
$\Sigma$ via the minimal sufficient parameters, $\gamma =(S_{vv},S_{v1},S_{11})'$, so that we will write
$m_{HJ}(\theta) = m_{HJ}(\theta, \gamma)$.

\subsubsection*{Example 2: Mean-Variance Analysis of Asset Portfolios}

Let us now describe the second problem. The classical \cite{M1952} problem is to minimize the variance of a portfolio given some attainable level of return:
$$
\min_w \Ep_t [ r_{p,t+1} - \Ep_t [r_{p,t+1}] ]^2\; \text{ such that } \; \Ep_t [r_{p, t+1}] = \mu,
$$
where $r_{p, t+1}$ is the return of the portfolio, determined as $r_{p, t+1}= w' r_{t+1}$, and $w$ is a vector of portfolio
``weights" and $r_{t+1}$ is a vector of returns on available assets. In a canonical version of the problem, we have that the vector of mean returns $v$ and covariance of returns $\Sigma$ do not vary with time period $t$, so that the problem becomes:
$$
\sigma_M(\mu) = \min_w w'\Sigma w \;\text{ such that }\; w'v = \mu.
$$
An explicit solution for $\sigma_M(\mu)$ takes the form,
$$
\sigma_M ^{2}\left( \mu \right) =\frac{S_{11}\mu ^{2}-2S_{v1}\mu +S_{vv}}{S_{vv}S_{11}-S_{v1}^{2}},
$$
where the minimal sufficient parameters $\gamma =(S_{vv},S_{v1}, S_{11})'$ are the same as in equation (\ref{define ABC}).

Therefore given a compact convex body $K\subset\mathbb{R}^2$, the Markowitz (M) set of admissible standard deviations and means in $K$ is given by
$$
\Theta_M = \{ (\mu, \sigma)
\in (\Bbb{R}\times\Bbb{R}_+) \cap K:
\sigma_M(\mu) - \sigma
\leq 0 \}=\{ \theta \in \Theta: m_M(\theta) \leq
0\},
$$
where $m_M(\theta):=\sigma_M(\mu)-\sigma$. The boundary of the set $\Theta_M$ is known as the efficient frontier. Note that as in HJ example, the inequality-generating function $m_M(\theta) = m_M(\theta, \gamma)$  depends on the unknown parameters, the means and covariance of returns, $\gamma =(S_{vv},S_{v1},S_{11})'.$  We also note that in some applications, the complementary Markowitz set is of interest:
$$
\Theta_{M}^c = \{ \theta \in \Theta: -m_M(\theta) \leq
0\}.
$$
The inference on such sets thus also fall in our framework (\ref{general}).


Mean-variance analysis for asset returns and the stochastic discount factor provides powerful tools to summarize the opportunities for risk diversification in a given asset market, and can serve as a basis for spanning and efficiency tests. The importance of sampling error and the role of statistical inference in portfolio analysis has long been recognized in the literature:
\cite{GRS1989} develop regression-based efficiency tests, and \cite{BJ1999} considers  inference on the tangency portfolio and tests for portfolio efficiency; recent work by \cite{PS2010} combines restrictions derived from the two dual approaches to mean-variance analysis in order to derive more powerful spanning tests for (sub-)markets; a summary of the recent literature on mean-variance efficiency tests is provided by \cite{S2009}.
The econometric contribution of this paper is quite distinct from and hence complementary to these previous efforts.

\subsubsection*{Example 3: Multi-Factor Efficient Frontiers}

A third example due to \cite{F1996} extends the unconditional portfolio choice problem in \cite{M1952} to multiple sources of priced risk. More specifically, suppose that there exist $k$ state variables $Z_t=(z_{1t},\dots,z_{kt})'$ (``factors") such that $(r_t',Z_t')'$ are jointly normal (or, more generally, jointly spherical). We assume that $Z_t$ follows a $K$-variate standard normal (or other spherical) distribution, and
\[\Ep_t[r_{t+1}|Z_{t+1}] = v + B'Z_{t+1}\]
where $v$ is an $N$-dimensional vector of average returns, and the $k\times N$ covariance matrix  $B=[\beta_1,\dots,\beta_N]=\textrm{Cov}_t(Z_{t+1},r_{t+1})$ gives the factor loadings for the vector of asset returns. As in the previous examples, we denote the (unconditional) mean of asset returns by $v\in\mathbb{R}^N$, and their covariance matrix with $\Sigma$. Note that for the purposes of this example, we take $r_{it}$ to denote the excess return of asset $i$ relative to a risk-free asset with return $r_f$.

This analysis is motivated by the investor's lifetime consumption-investment problem, where the state variables $Z_t$ denote other factors influencing prices for consumption goods and the household's income from wage labor or entrepreneurial activity in period $t$. In particular the household not only faces a trade-off between mean and variance of portfolio returns, but may also wish to use her portfolio to insure against price or income shocks.

Specifically, we consider the problem constructing a minimum variance portfolio that targets a mean return $\mu^*$ and a vector $\beta^*$ of loadings respect to the factors $Z_t$ in the absence of borrowing constraints. Denoting $D:=[v,B']$ and $\delta^*:=(\mu^*,\beta^{*'})'$, the efficient portfolio weights $w$ can be found by solving the program
\begin{equation}
\min_w w'\Sigma w\hspace{0.5cm}\textnormal{s.t. }D'w=\delta^*.
\end{equation}
Note that since the investor has access to the risk-free asset and faces no short-sale constraints, the portfolio weights $w$ are not required to be nonnegative or sum up to one. Solving this program yields efficient portfolio weights $w^*=\Sigma^{-1}D(D'\Sigma D)^{-1}\delta^*$ so that the lower bound on the variance of a portfolio with beta equal to $\beta$ and mean $\mu$ is given by
\[\sigma_{MF}^2(\mu,\beta):=(w^*)'\Sigma w^*=\delta^{*'}(D'\Sigma D)^{-1}\delta^*.\]

The resulting multi-factor efficient frontier bounds the cone
\[\Theta_{MF}:=\left\{(\mu^*,\beta^{*'},\sigma)'\in(\mathbb{R}^{k+1}\times\mathbb{R}_+)\cap K:\sigma_{MF}(\mu^*,\beta^*)-\sigma\leq0\right\}\]
of feasible mean-variance-beta combinations, where $K$ is some compact convex body in $\mathbb{R}^{k+2}$. Again, this cone can also be characterized by an inequality condition on the moment function
\[m_{MF}(\theta,\gamma):=\sqrt{(\mu^*,\beta^{*'})(D'\Sigma D)^{-1}(\mu^*,\beta^{*'})'}-\sigma\]
where $\theta=(\mu^*,\beta^{*'},\sigma)'$ and $\gamma = \textnormal{vec}(D'\Sigma D)$.

Extending the two-fund theorem, \cite{F1996} shows that we can always construct $k+2$ different funds from the $N$ assets that span the entire multi-factor efficient frontier. In the presence of heterogeneity in preferences, wealth and consumption possibilities, different investors will typically choose different points on the efficient cone, so that only the entire set constitutes an adequate representation of the opportunities to insure against different factor risks in the market.

\subsubsection*{Example 4: Bounds on Demand Elasticities}
To illustrate the wider applicability of our methods, we also
consider inference on demand elasticities in the presence of an
optimization friction. \cite{C2012} considers a household's dynamic
labor supply and consumption problem and proposes bounds on price
elasticities that allow for optimization errors of unknown form but
that are bounded in magnitude. His analysis assumes that preferences
can be represented by quasilinear flow utilities and lead to
Marshallian demand functions of the form
\[\log x_{it}^*(p) = \alpha - \varepsilon\log p +\nu_{it},\]
where $p$ is the price of the good of interest, and $\nu_{it}$ a
preference shock for household $i$ in period $t$. Here $\varepsilon$
is the (constant) ``structural" demand elasticity corresponding to the
first-best solution to the household's planning problem.\\

Let $U_{it}^*$ be the discounted lifetime utility from following the
optimal consumption plan and let $e_{it}(x)$ denote the minimal
expenditure needed to finance a utility level $U_{it}^*$ under the
constraint that $x_{it}$ is held fixed at $x$. The analysis then
considers ``small" optimization errors in the households' observed
choice of $x_{it}$ such that for every period $t$, the average utility
loss measured in terms of the difference in expenditure as a
percentage of the optimal budget is bounded by some $\delta>0$, i.e.
\[\frac1N\ssum_{i=1}^N\frac{e_{it}(x_{it}) -
e_{it}(x_{it}^*(p_t))}{p_tx_{it}^*}\leq\delta,\]
where the average is taken over a sample of households $i=1,\dots,N$.

\cite{C2012} derives the lower and upper bounds $\varepsilon_L$ and $\varepsilon_U$ for the structural elasticities that are compatible an observed elasticity $\varepsilon^o$ given a price change $\Delta\log p$, and in the presence of an optimization error of $\delta>0$. This bound is given by the following two inequalities
\[\varepsilon^o = \varepsilon_U - 2\frac{(2\varepsilon_U\delta)^{1/2}}{\Delta\log p}\hspace{0.5cm}\textnormal{and }\varepsilon^o = \varepsilon_L + 2\frac{(2\varepsilon_L\delta)^{1/2}}{\Delta\log p}\]
where $\delta$ is assumed to be small.\footnote{See equations (12) and (13) on pages 983-985 in \cite{C2012}} With a few basic manipulations, we can obtain the following equivalent characterization of the $(\delta,\varepsilon)$-set as a lower bound for the distortion needed to reconcile $\varepsilon^o$ with a structural elasticity $\varepsilon$:
\begin{equation}\label{chetty_bound}
\delta\geq \frac{(\varepsilon-\varepsilon^o)^2(\Delta\log p)^2}{8\varepsilon}=:\delta_{OF}(\varepsilon,\varepsilon^o,\Delta\log p)
\hspace{0.5cm}\textnormal{if and only if  }\varepsilon\in[\varepsilon_L,\varepsilon_U].
\end{equation}
\cite{C2012} gives several interpretations for this optimization friction, which could result e.g. from adjustment costs or misperception of prices. In the context of the consumption-CAPM model, \cite{LL2009} conducted a similar sensitivity analysis, allowing for optimization errors in terms of the Euler equations characterizing the household's optimal investment decision.

The resulting set of structural elasticity/optimization friction pairs $(\varepsilon, \delta)$ compatible with the observed elasticity $\varepsilon^o$ is given by
\[\Theta_{OF}:=\left\{(\varepsilon,\delta)\in\mathbb{R}_+^2:\delta_{OF}(\varepsilon,\varepsilon^o,\Delta\log p)-\delta\leq0\right\}.\]
The bounds in equation (\ref{chetty_bound}) can therefore also be expressed as an inequality condition on the moment function
\[m_{OF}(\theta,\gamma):=\frac{(\varepsilon-\varepsilon^o)^2(\Delta\log p)^2}{8\varepsilon}-\delta,\]
where $\theta = (\varepsilon,\delta)$ and  $\gamma = \varepsilon^o$, and $\Delta\log p$ will be regarded as a fixed design parameter.
In applications, one might be interested in intersections bounds from several studies in order to obtain tighter bounds, as was done in \cite{C2012}.
As we discuss below and illustrate in the empirical example, it is possible to perform this intersection in a smooth fashion, retaining the general single-equation structure as given in (\ref{general}).

\subsection{Overview}

The basic problem of this paper is to develop inference methods on sets defined by nonlinear inequality restrictions, as the HJ and M sets, while accounting for uncertainty in the estimation of parameters of the inequality-generating functions. Specifically, the problem is to construct a confidence region $R$ such that
$$
\lim_{n \to \infty} \Pp \{ \Theta_0 \subseteq R \}
= 1- \alpha, \ \    d_H(\Theta_0, R) = O_p(1/\sqrt{n}),
$$
where $d_H$ is the Hausdorff distance, defined below in (\ref{eq: H-distance}).  We will construct confidence regions for the set $\Theta_0$ using LR and Wald-type Statistics, building on and simultaneously
enriching the approaches suggested in \cite{CHT2007} and in \cite{BM2008} and \cite{M1997}, respectively.
We also would like to ensure that confidence regions $R$ are as small as possible and converge to $\Theta_0$ at the fastest attainable speed.

Once $R$ is constructed, we can test any composite hypotheses involving the parameter $\theta$ without compromising the significance level. E.g. a typical application of the HJ bounds determines which combinations of $\Theta_e$ the first two moments of the SDF generated by a given family of economic models fall in the HJ set.  Indeed, in that case $\Theta_e = \{ \mu(\varrho), \sigma(\varrho), \varrho \in [0,\infty)\}$, where $\varrho$ is the elasticity of power utility function, and we can check which values of $\gamma$ give us overlap of $\Theta_e$ with the confidence set $R$ for the HJ set.\footnote{In cases where curve $\varrho \mapsto \mu(\varrho), \varrho\mapsto \sigma(\varrho)$ are estimated, we need
to construct a confidence set $R_e$ for $\Theta_e$ and then look whether (and where) $R_e$ and $R$ overlap to determine the plausible values of $\varrho$. }  Similar comments about applicability of our approach go through for the M and MF sets as well. We should also mention
that our confidence regions $R$ are also more powerful  than the structured projection methods, which rely upon building confidence sets for the frontier-determining sufficient parameters (e.g. frontier-spanning portfolios), and then projecting them to obtain confidence sets for HJ sets or MF sets.  We demonstrate this in the empirical section, where we compute the confidence sets for HJ sets based on our approach and based on structured projection approach, and show that our confidence interval is much smaller and lies strictly inside the confidence region based on structured projections.  We provide a theoretical explanation to this phenomenon in Section 2.5

Our procedure for inference using weighted Wald-type statistics complements other approaches based on the directed Hausdorff distance suggested in \cite{BM2008} and Hausdorff distance in \cite{M1997}. By using weighting in the construction of the Wald-type statistics, we make this approach asymptotically invariant to parameter transformations, which results in noticeably sharper confidence sets, at least in the canonical empirical example that we will show. These invariance  properties should be seen as complementary to questions of efficient estimation of the identified set as studied by \cite{KS2011}; indeed, many types of confidence sets will all be centered around the same efficient set estimate, and so we use invariance and precision considerations to select amongst various potential constructions of sets.

\cite{CMS1996} analyzed the geometric structure of Wald-type tests with nonlinear restrictions for point-identified problems and pointed out the beneficial role of invariance of the distribution of the statistic with respect to reparameterizations. In regular problems, the effect of nonlinearities on inference is typically asymptotically negligible due to the delta method and its extensions. However, our results show that for inference on sets with a nontrivial diameter, the effect of nonlinearities is of the same order as that of sampling variation. In particular we propose inference procedures that rely on quantities that are asymptotically pivotal and (asymptotically) invariant to parameter transformation, and illustrate that procedures failing to meet these requirements lead to overly conservative inference. In particular, these conditions are not met by the classical Hausdorff distance except in very special cases.

Thus, our construction is of independent interest for this type of inference and is a useful complement to the work
of \cite{BM2008} and \cite{M1997}. Furthermore, our results on formal validity of the bootstrap for the weighed LR-type and W-type statistics are also of independent interest (for the class of problems with structures that are similar to those studied here; see \cite{BM2008}
and \cite{KS2011} for related bootstrap results for Hausdorff statistics.)

The rest of the paper is organized as follows.  In Section 2 we present our estimation and inference results.  In Section 3 we present an empirical example, illustrating the constructions of confidence sets for HJ sets.  In Section 4 we draw conclusions and provide direction for further research. In the Appendix, we collect the proofs of the main results.

\section{Estimation and Inference Results}

\subsection{Basic Constructions} We first introduce our basic framework.
We have a real-valued inequality-generating function $m(\theta)$, and the set of interest is the solution of the inequalities generated by the function $m(\theta)$ over a compact parameter space $\Theta\subset\mathbb{R}^k$:
$$
\Theta_0 = \{ \theta \in \Theta: m(\theta) \leq
0\}.
$$
A natural estimator of $\Theta_0$ is its empirical analog
$$
\widehat \Theta_0 = \{ \theta \in \Theta: \widehat m
(\theta) \leq 0 \},
$$
where  $\widehat m
(\theta)$ is the estimate of the inequality-generating function.
For example, in the HJ and M examples, the estimate takes the form
$$
\widehat m(\theta) = m(\theta, \widehat \gamma), \ \ \widehat
\gamma =   (\widehat S_{vv}, \widehat S_{v1},\widehat S_{11} )'.
$$

\begin{remark}[Approximating Multiple Inequalities by A Single Smooth Inequality] Throughout the paper, we will only consider the case of a single inequality restriction, $m:\Theta\rightarrow\mathbb{R}$ satisfying certain smoothness conditions with respect to $\theta$. However this framework approximately encompasses the case of multiple inequalities,
since it is possible to conservatively approximate intersection bounds of the form $g(\theta)\leq 0$, where $g(\theta):=(g_1(\theta),\dots,g_J(\theta))'$ represents a vector of constraints, by a single smooth bound $m (\theta) \leq 0$.  To be specific, we can form an inequality-generating function:
\[m_{\lambda}(\theta) := m(g_1(\theta),\dots,g_J(\theta);\lambda):=\ssum_{j=1}^J\frac{\exp(\lambda g_j(\theta))}{\ssum_{l=1}^J
\exp(\lambda g_l(\theta))}g_j(\theta),\]
where $\lambda\in\mathbb{R}$ is a fixed, positive scalar.  This function $m_{\lambda}$ is clearly smooth with respect to the underlying function $g$, and is a conservative approximation to $\max_{1 \leq j \leq J} g_j(\theta)$ with an explicit error bound, as shown by the following lemma.

\begin{lemma}[Properties of Smooth Max] \label{smooth_appx_bd} (i) We have that
$$
m (g_1,\dots,g_J,\lambda) \leq 0 \ \text{ if } \ \max_{1 \leq j \leq J} g_j(\theta) \leq 0.
$$
(ii) For any $\lambda>0$, the approximation error obeys
\[\sup_{(g_1,\dots,g_J)'\in\mathbb{R}^J}\left|\max(g_1,\dots,g_J) - m (g_1,\dots,g_J,\lambda)\right|
\leq \lambda^{-1} W\left(\frac{J-1}e\right)\]
where $e$ is Euler's number, and the function $W:\Bbb{R} \mapsto \Bbb{R}$ is Lambert's product logarithm function, i.e. $W(z)$ is defined as the solution $w$ of the equation $w\exp\{w\}=z$. In particular, $W(z)\geq0$ if $z \geq 0$, and $W(z)\leq \log(z)$ if $z\geq e$.
\end{lemma}
This lemma implies that the approximation error relative to the max function (with respect to the sup-norm) is inversely proportional to the smoothing parameter $\lambda$, where the constant of proportionality grows only very slowly in $J$.
In the paper we will state our results and assumptions on estimation and inference directly in terms of the scalar moment $m(\theta)= m_{\lambda}(\theta)$ and its empirical analog $\widehat m(\theta):= \widehat{m}_{\lambda}(\theta):=m(\widehat{g}_1(\theta),\dots,\widehat{g}_J(\theta);\lambda)$.
We also consider the quantity $1/\lambda$ fixed at some small value.
\qed
\end{remark}

Our proposals for confidence regions are based on (1) a LR-type statistic and (2)  a Wald-type
statistic. The LR-based confidence region is
 \begin{equation}
R_{LR}:= \left \{ \theta \in \Theta: \left[\sqrt{n} \widehat
m(\theta)/\widehat{s}(\theta) \right ]^2_+ \leq \widehat k(1-\alpha) \right
\},
 \end{equation}
where $\widehat{s}(\theta)$ is a weighting function;
ideally, the standard error of $\widehat m(\theta)$;
and $\widehat k(1-\alpha)$ is some suitable estimate of
$k(1-\alpha)$, the  $(1- \alpha)$-quantile of the statistic
 \begin{equation}
\mathcal{L}_n = \sup_{\theta \in \Theta_0}
\left[\sqrt{n} \widehat m(\theta)/\widehat{s}(\theta) \right ]^2_+.
 \end{equation}
Note that $\mathcal{L}_n$ is a LR-type statistic, as in \cite{CHT2007}.

Next we shall consider confidence regions based on inverting a Wald-type statistic. In order to eliminate boundary effects on the distribution of
this statistic, we will need to assume that the moment conditions, and therefore the set estimator, are well-defined in a neighborhood of the parameter space.  To this end, we define the distance of a point $\theta$ to a set $A\subset\mathbb{R}^k$ as
$$ d(\theta, A)
:= \inf_{\theta' \in A} \| \theta
- \theta' \|. $$ We let the set $$\Theta^{\delta}:=\{\theta\in\mathbb{R}^k:d(\theta,\Theta)\leq\delta\}$$ denote the $\delta$-expansion of $\Theta$ in $\mathbb{R}^k$ for any $\delta>0$. We denote the natural set estimator for $\Theta_0$ in the expansion of the parameter space by
\[\widehat{\Theta}_{0,\delta}:=\left\{\theta\in\Theta^{\delta}:\widehat{m}(\theta)\leq0\right\}.\]
We will see below that the behavior of the moment function near the boundary of the parameter space is in general relevant for the statistical behavior of our procedure. However, for the practical applications discussed in this paper, the restriction of the parameter space $\Theta$ to a compact set is entirely for technical reasons, so that an extension of the definitions to $\Theta^{\delta}$ is unproblematic.

Given these definitions, our Wald-based confidence region is
 \begin{equation}
R_W := \{ \theta \in \Theta: [\sqrt{n} d (\theta, \widehat{\Theta}_{0,\delta})/\widehat{w}(\theta)]^2 \leq \widehat k(1-\alpha) \},
 \end{equation}
where $\widehat{w}(\theta)$ is a weighting function, particular forms
of which we will suggest later;
and $\widehat k(1-\alpha)$ is a suitable estimate of
$k (1-\alpha)$, the $(1-\alpha)$-quantile of
$\mathcal{W}_n,$ where  $\mathcal{W}_n$ is the weighted W-statistic
\begin{equation}\label{statw}
\mathcal{W}_n :=  \sup_{\theta \in \Theta_0} [\sqrt{n} d
(\theta, \widehat{\Theta}_{0,\delta})/\widehat{w}(\theta)]^2.
 \end{equation}
In the special case, where the weight
function is flat, namely $\widehat{w}(\theta) = w$ for all $\theta$, the W-statistic $\mathcal{W}_n$ becomes the
canonical directed Hausdorff distance (see \cite{M1997}, and \cite{BM2008}):
$$
\sqrt{\mathcal{W}_n} \propto   d^+(\Theta_0, \widehat
\Theta_{0,\delta}) = \sup_{\theta \in \Theta_0}
\inf_{\theta' \in \widehat{\Theta}_{0,\delta} } \| \theta -
\theta' \|.
$$

\begin{remark}[Invariance Motivation for Weighed W-Statistics] The weighted W-statistic (\ref{statw}) is generally not a distance, but we argue in Section 2.6 that it provides a very useful extension/generalization of the canonical directed Hausdorff distance.  Note that, like any Euclidean norm, the Hausdorff distance is not invariant with respect to changes in the scale of the different components of $\theta$, which results in confidence regions that are not equivariant to such transformations. Our empirical results below illustrate that as a result, the shape of confidence sets alters dramatically as we change the weight on different dimensions of the parameter space. In sharp contrast, the introduction of certain types of weights makes
the W-statistic (first-order) invariant to such transformations, as shown in Section 2.6, which makes the resulting confidence regions (first-order) equivariant. As a result, in all of our empirical examples such weighting dramatically improves the confidence regions (e.g. compare Figures
\ref{weightedhausdorffdistancewithbootstrappedcurves} and \ref{hausdorffdistancewithbootstrappedcurves} in the example
concerning inference on HJ sets.) \qed
\end{remark}

\subsection{A Basic Limit Theorem for LR and W statistics}

In this subsection, we develop a basic result on the limit laws of the LR and W statistics. We will develop this result under the following general regularity conditions:\\[8pt]

\textsc{ Condition R.}\textit{ For some $\delta>0$, the inequality-generating functions $m:\Theta^{\delta}\rightarrow\mathbb{R}$ and its estimator $\widehat{m}:\Theta^{\delta}\rightarrow\mathbb{R}$ are well-defined.
\begin{itemize}
\item[(R.1)] The estimator $\widehat m$ is $n^{-1/2}$-consistent for $m$ and asymptotically Gaussian, namely, in the metric space of bounded functions $\ell^{\infty}( \Theta^{\delta})$
    $$
   \sqrt{n} (\widehat m - m) \rightsquigarrow  \Gp ,
    $$
    where $\Gp$ is a zero-mean Gaussian process with continuous paths a.s., and a non-degenerate covariance function, i.e. $\inf_{\theta \in \Theta^{\delta}} \Ep[\Gp^2(\theta)] >0$.
\item[(R.2)] Functions $\widehat m$ and $m$ admit  continuous gradients $\nabla_\theta \widehat m:\Theta^{\delta} \mapsto \Bbb{R}^k$ and $\nabla_\theta  m:\Theta^{\delta} \mapsto \Bbb{R}^k$ with probability one, where
    $$
   \nabla_\theta \widehat m(\theta) = \nabla_\theta m(\theta) + o_p(1).
    $$
    uniformly in $\theta \in \Theta^{\delta}$. The gradient $\theta \mapsto \nabla_\theta m$ is uniformly Lipschitz and $\inf_{\theta \in \Theta^{\delta}}\|\nabla_\theta m(\theta)\|>0.$
    \item[(R.3)]  Weighting functions $\widehat s: \Theta^{\delta} \mapsto \Bbb{R} $ and $s: \Theta^{\delta} \mapsto \Bbb{R} $ satisfy uniformly in $\theta \in \Theta^{\delta}$
$$
\widehat{s}(\theta) = s(\theta)  + o_p(1), \
\  \widehat{w}(\theta) = w(\theta)  + o_p(1),
$$
where
$s: \Theta^{\delta} \mapsto \Bbb{R}_+$ and $w: \Theta^{\delta} \mapsto \Bbb{R}_+$ are continuous functions with values
bounded away
from zero.
\end{itemize}
}
In Condition R.1, we require the estimates of the
inequality-generating functions to satisfy a uniform central limit
theorem. Many sufficient conditions for this are
provided by the theory of empirical processes, see e.g. \cite{vaart}. In our finance examples, this
condition will follow from asymptotic normality of the estimates of
the mean returns and covariance of returns.  In Condition R.2, we
require that gradient of the estimate of the inequality-generating
function be consistent for the gradient of the inequality-generating
function.  Moreover, we require that the norm
$\|\nabla_\theta m(\theta)\|$ be bounded away from zero, which is an identification condition and allows us to
estimate, at a parametric rate, the boundary of the set $\Theta_0$,
which we  define as
$$
\partial \Theta_0 := \{ \theta \in \Theta : m(\theta) = 0\}.
$$
In Condition R.3, we require that the estimates of the weight functions be consistent, and the weight functions be well-behaved.

Under these conditions we can state the following general result.

\begin{theorem}\label{theorem1}\textbf{(Limit Laws of LR and W Statistics).}  Under Condition R
\begin{eqnarray}
\mathcal{L}_n &\rightsquigarrow&   \mathcal{L}, \ \ \ \mathcal{L} = \sup_{\theta
\in \partial \Theta_0} \left [\frac{\Gp(\theta)}{s(\theta) }\right]^2_+, \\
\mathcal{W}_n & \rightsquigarrow&   \mathcal{W}, \ \ \mathcal{W} = \sup_{\theta\in
\partial \Theta_0} \left [ \frac{\Gp(\theta) }{\|
\nabla_{\theta} m(\theta) \| \cdot
w(\theta) }  \right]^2_+,
 \end{eqnarray}
where both $\mathcal{W}$ and $\mathcal{L}$ have distribution functions that
are continuous at their $(1-\alpha)$-quantiles for $\alpha < 1/2$.  Furthermore, if
$$
\widehat{w}(\theta) = \frac{s(\theta)}{\|
\nabla_{\theta} m(\theta) \| } + o_p(1),
$$
uniformly in $\theta \in \Theta^{\delta}$, then the two statistics are asymptotically equivalent:
$$
\mathcal{W}_n = \mathcal{L}_n + o_p(1).$$
In particular, this equivalence occurs if $\widehat{w}(\theta) = \|
\nabla_{\theta} \widehat m(\theta) \|/\widehat s(\theta)$.
\end{theorem}
We see from this theorem that the LR and W statistics converge in law to well-behaved random variables that are continuous transformations of the limit Gaussian process $\Gp$.  Moreover, we see that under an appropriate choice of the weighting functions, the two statistics are asymptotically equivalent.

For our application to HJ and MF sets, the following conditions will be sufficient. \\

\textit{\textsc{Condition C.}
\begin{itemize}
\item[(C.1)] We have that $m(\theta) = m(\theta, \gamma_0)$, where $\gamma_0 \in \Gamma \subset \Bbb{R}^d$, for all $\theta \in \Theta^{\delta}$ and some $\delta>0$.  The value $\gamma_0$ is in the interior of $\Gamma$, and there is an estimator $\widehat \gamma$ of  $\gamma_0$ that obeys $$\sqrt{n} (\widehat \gamma -
\gamma_0) \rightsquigarrow \Omega^{1/2} Z,  \ \ \ Z = N(0,
I_d).$$
for some positive-definite Hermitian matrix $\Omega$.
Moreover, there is $\widehat \Omega \to_p \Omega$.
\item[(C.2)]  The gradient map $(\theta, \gamma) \mapsto \nabla_{\theta} m(\theta,
\gamma)$, mapping $\Theta^{\delta} \times \Gamma \mapsto \Bbb{R}^k$, exists and is uniformly Lipschitz-continuous. Moreover,
$\inf_{(\theta,\gamma) \in \Theta^{\delta} \times \Gamma}\|\nabla_{\theta} m(\theta,\gamma)\|>0$.
\item[(C.3)] The gradient map $(\theta, \gamma)\mapsto \nabla_{\gamma} m(\theta, \gamma)$, mapping $\Theta^{\delta} \times \Gamma \mapsto \Bbb{R}^k$, exists and is uniformly Lipschitz-continuous.
\end{itemize}}
We show in Proposition \ref{hj_m_c1_2_prp} that these conditions hold for the canonical versions
of the HJ and MF problems. Under these conditions we immediately conclude that in the metric space of bounded functions $\ell^{\infty}(\Theta^{\delta})$:
 \begin{eqnarray}
 \nonumber\sqrt{n} (\widehat m(\cdot) - m(\cdot)) & =&  \nabla_{\gamma} m (\cdot,\bar \gamma)'
 \sqrt{n}(\widehat \gamma - \gamma_0) + o_p(1) \\
 \nonumber&\rightsquigarrow& \nabla_{\gamma} m (\cdot,\gamma_0)'
 \Omega^{1/2} Z,
 \end{eqnarray}
where $\nabla_{\gamma} m (\theta,\bar \gamma)$ denotes the gradient with each of its rows evaluated
at a  value $\bar \gamma$ on the line connecting $\widehat \gamma$ and $\gamma_0$,
 where value $\bar \gamma$ may vary from row to row of the matrix.  Therefore, the limit
 process in HJ and M examples takes the form:
 \begin{equation}\label{cor1-1}
 \Gp(\theta) = \nabla_{\gamma} m (\theta,\gamma_0)'
 \Omega^{1/2} Z.
 \end{equation}
 This will lead us to conclude formally below that conclusions of Theorem 1 hold with
\begin{eqnarray}\label{cor1-2a}
 \mathcal{L} & = & \sup_{\theta
\in \partial \Theta_0} \left [
\frac{\nabla_{\gamma} m(\theta, \gamma_0)'
\Omega^{1/2}}{s(\theta) }  Z \right]^2_+, \\
\mathcal{W} & = & \sup_{\theta
\in
\partial \Theta_0} \left [ \frac{\nabla_{\gamma}
m(\theta, \gamma_0)' \Omega^{1/2} }{\|
\nabla_{\theta} m(\theta, \gamma_0) \| \cdot
w(\theta) }Z \right]^2_+. \label{cor1-2b}
\end{eqnarray}

A good strategy for choosing the weighting function for LR and W is to choose the studentizing
Anderson-Darling weights
 \begin{eqnarray}\label{cor1-3a}
s(\theta) & = &  \|\nabla_{\gamma} m (\theta,\gamma_0)'
 \Omega^{1/2}\|, \\
w(\theta) & = & \frac{\| \nabla_{\gamma}
m(\theta, \gamma_0)' \Omega^{1/2} \|}{ \|
\nabla_{\theta} m(\theta, \gamma_0) \| }. \label{cor1-3b}
 \end{eqnarray}
The natural estimates of these weighting functions are given by the following plug-in estimators:
 \begin{eqnarray}
 \label{cor1-4a}
\widehat{s}(\theta) := \|\nabla_{\gamma} m (\theta,\widehat \gamma)'
 \widehat \Omega^{1/2}\|, \\
  \widehat{w}(\theta) := \frac{\| \nabla_{\gamma}
m(\theta, \widehat \gamma)' \widehat \Omega^{1/2} \|}{ \|
\nabla_{\theta} m(\theta, \widehat \gamma) \| }. \label{cor1-4b}
 \end{eqnarray}

We formalize the preceding discussion as the following corollary. \\

\begin{corollary}\label{corollary1}\textbf{(Limit Laws of LR and W statistics under Condition C).}
Under Condition C, Conditions R holds with the limit Gaussian process stated in equation (\ref{cor1-1}). The plug-in estimates of the weighting functions (\ref{cor1-4a}) and (\ref{cor1-4b}) are uniformly consistent for the weighting functions (\ref{cor1-3a}) and (\ref{cor1-3b}). Therefore,
conclusions of Theorem 1 hold with the limit laws for our statistics given by the laws
of random variables stated in equations (\ref{cor1-2a}) and (\ref{cor1-2b}).
\end{corollary}

\subsection{Basic Validity and Convergence Rates for the Confidence Regions}

We will first give a basic validity result for confidence regions assuming that we have suitable estimates of the quantiles of LR and W statistics and will verify basic validity of our confidence regions. A basic procedure for constructing suitable estimates of these quantiles via bootstrap or simulation will be given below.

\begin{theorem}\label{theorem2}\textbf{(Basic Inferential Validity of Confidence Regions).} Suppose that for $\alpha < 1/2$ we have consistent estimates of quantiles of limit statistics $\mathcal{W}$ and $\mathcal{L}$, namely,
\begin{equation}\label{quantile consistency}
\widehat k(1-\alpha) = k(1-\alpha) + o_p(1),
\end{equation}  where $k (1-\alpha)$ is the $(1-\alpha)$-quantile
of either $\mathcal{W}$ or $\mathcal{L}$, respectively.  Then  as the sample size $n$ grows to infinity, confidence regions $R_{LR}$  and $R_W$ cover $\Theta_0$ with probability approaching $1-\alpha$:
\begin{eqnarray}
&& \Pp[ \Theta_0 \subseteq R_{LR} ] =  \Pp[ \mathcal{L}_n \leq \widehat k(1-\alpha) ] \to \Pp[ \mathcal{L} \leq k(1-\alpha) ] = (1-\alpha), \\
&& \Pp[ \Theta_0 \subseteq R_{W} ] \ =  \Pp[ \mathcal{W}_n \leq \widehat k(1-\alpha) ] \!\to \Pp[ \mathcal{W} \leq k(1-\alpha) ] \! = \! (1-\alpha).
\end{eqnarray}\end{theorem}

We next recall that given a Euclidian metric $d(\cdot,\cdot)$ on $\Bbb{R}^K$, the (symmetric) Hausdorff distance between two non-empty sets $A,B\subset\Bbb{R}^k$ is defined as
\begin{equation}\label{eq: H-distance}
d_H(A,B):= \max\left\{\sup_{a\in A}d(a,B),\sup_{b\in B}(b,A)\right\}.
 \end{equation}
Our next result shows that the confidence regions based on the LR and the Wald statistic are also root-n consistent estimators for the set $\Theta_0$
with respect to the Hausdorff distance. This result also establishes that the Pitman rates for $R_W$ and $R_{LR}$ is also root-$n$, namely any alternative set $\Theta_A$ such that $\sqrt{n}d_H(\Theta_0, \Theta_A) \to \infty$ will not be covered by $R_W$ and $R_{LR}$, and hence will be rejected with probability tending to 1.

\begin{theorem}\label{root_n_consistency}\textbf{(Confidence Regions are $\sqrt{n}$-Consistent Estimator of $\Theta_0$).} Under Condition R, the confidence regions based on the LR and Wald statistics are consistent with respect to the Hausdorff distance at a root-n rate, that is
$$d_H(R_{LR},\Theta_0)=O_p(n^{-1/2}) \text{ and } d_H(R_{W},\Theta_0)=O_p(n^{-1/2}).$$
As a consequence, if $\sqrt{n} d_H(\Theta_0,\Theta_A) \to \infty$, then
$$\Pp[ \Theta_A \subseteq R_{LR}] \to 0 \text{ and } \Pp[ \Theta_A \subseteq R_W] \to 0.$$
\end{theorem}

While the results in Theorems \ref{theorem1}-\ref{root_n_consistency} were stated in terms of the high-level Condition R, the next corollary shows that these results can be applied to any problem where Condition C holds. These conditions hold in
HJ, MF, and other problems listed in Section 1, as we verify below.\\

\begin{corollary}\label{corollary2}\textbf{(Limit Laws of LR and W statistics under Condition C).}
Suppose that Condition C holds and that consistent estimates of quantiles
of statistics (\ref{cor1-2a}) and (\ref{cor1-2b}) are available.  Then the conclusions
of Theorem 2 apply.  \end{corollary}


We now turn to estimation of quantiles for the LR and W statistics using bootstrap, simulation, and other resampling schemes under general conditions.  The basic idea is as follows: First, let us take any procedure that consistently estimates the law of our basic Gaussian process $\Gp$
or a weighted version of this process appearing in the limit expressions.  Next, we can use the estimated law to obtain consistent estimates of the laws of the LR and W statistics, and thus also obtain consistent estimates of their quantiles. It is well known that there are many
procedures for accomplishing the first step, including such common schemes as the bootstrap, simulation, and subsampling, which can also be adapted to allow for various forms of cross-section and time series dependence.

In what follows, we will simplify the notation by writing our limit
statistics as a special case of the following statistic:
\begin{equation}
\mathcal{S} := \sup_{\theta \in \partial \Theta_0} [ V(\theta)]^2_+, \ \ \ V(\theta) := \tau (\theta) \Gp(\theta).
\end{equation}
Thus,  $\mathcal{S}= \mathcal{L}$ for $\tau(\theta) = 1/s(\theta)$ and
$\mathcal{S} = \mathcal{W}$ for $\tau(\theta) = 1/[\|\nabla_{\theta} m(\theta)\| \cdot w(\theta)]$.   We take $\tau$ to be a continuous function bounded away from zero on the parameter space.
We also need to introduce the following notations and concepts.   Our process
$V$ is a random element that takes values in the metric space of continuous functions $C(\Theta)$ equipped with the uniform metric. The underlying measure space is $(\Omega, \mathcal{F})$ and we denote the law of $V$ under the probability measure $P$ by the symbol $\mathcal{Q}_V$.

In the following we will assume that we have an estimate $\mathcal{Q}_{V^*}$ of  the law $\mathcal{Q}_V$ of the Gaussian process $V$. This estimate $\mathcal{Q}_{V^*}$ is a probability measure which can be generated as follows:   Let us fix another measure space $(\Omega', \mathcal{F}')$ and a probability measure $P^*$ on this space. Then given a random element $V^*$ on this space taking values in $C(\Theta)$, we denote its law under $P^*$ by $\mathcal{Q}_{V^*}$. We thus identify the probability measure $P^*$ with a data-generating process by which we generate draws or realizations of $V^*$.  This identification allows us to cover such methods of producing realizations of  $V^*$ as the bootstrap, subsampling, or other simulation approaches.

We require that the estimate $\mathcal{Q}_{V^*}$ be consistent for $\mathcal{Q}_V$ in any metric $\rho_{K}$ metrizing weak convergence, where we can take the metric to be the Kantarovich-Rubinstein metric. Note that there are many results that verify this basic consistency condition for different processes $V$ and various bootstrap, simulation, and subsampling schemes, as we will discuss in more detail below.

To define the  Kantarovich-Rubinstein metric, let $\theta \mapsto
v(\theta)$ be an element of a metric space $(M,d_M)$, and
$\mathrm{BL}_1(M)$  be a class of Lipschitz functions $\varphi:
M \rightarrow \Bbb{R} $ that satisfy:
$$
|\varphi(v) - \varphi(v')| \leq d_M(v,v'), \ \ \ \sup_{v \in M}|\varphi(v) | \leq 1.
$$
The Kantarovich-Rubinstein distance  between probability laws $\mathcal{Q}_V$ and $\mathcal{Q}'_V$
of random elements $V$ and $V'$ taking values in $M$ is defined as:
 $$\rho_{K}(\mathcal{Q}, \mathcal{Q}'; M) := \sup_{\varphi \in \mathrm{BL}_1(M)}
|\mathrm{E}_{\mathcal{Q}} \varphi(V) - \mathrm{E}_{\mathcal{Q}'} \varphi(V) |.$$
As stated earlier, we require the estimate $\mathcal{Q}_{V^*}$ to be consistent for $\mathcal{Q}_V$
in the metric $\rho_{K}$, that is
 \begin{equation}\label{boot consistency}
\rho_{K}( \mathcal{Q}_{V^*}, \mathcal{Q}_V; C(\Theta)) = o_p(1).
 \end{equation}

Let $\mathcal{Q}_{\mathcal{S}}$ denote the probability
law of $\mathcal{S} = \mathcal{W}$ or $\mathcal{L}$, which is in turn induced by the law $\mathcal{Q}_V$ of the Gaussian process $V$.  We need to define the estimate $\mathcal{Q}_{\mathcal{S}^*}$ of this law. First,
we define the following plug-in estimate of the boundary set  $\partial \Theta_0$,
 \begin{equation}
\widehat{\partial \Theta_0} = \{ \theta \in \Theta: \widehat m(\theta) = 0 \}.
 \end{equation}
This estimate turns out to be consistent at the usual root-$n$ rate, by an argument similar to that given in \cite{CHT2007}.
Next, define $\mathcal{Q}_{\mathcal{S}^*}$  as the law of the random variable
\begin{eqnarray} \label{define law 1}
\mathcal{S}^* = \sup_{\theta
\in \widehat{\partial \Theta_0}} [V^*(\theta)]^2_+ \label{define law 2}.
 \end{eqnarray}
 In this definition, we hold the hatted quantities fixed, and the only random element is $V^*$ that is drawn according to the law   $\mathcal{Q}_{V^*}$.

We will show that the estimated law $\mathcal{Q}_{\mathcal{S}^*}$ is consistent for $\mathcal{Q}_{\mathcal{S}}$  in the sense that
 \begin{equation}\label{boot consistency 2}
\rho_{K}( \mathcal{Q}_{\mathcal{S}^*}, \mathcal{Q}_{\mathcal{S}}; \Bbb{R}) = o_p(1).
 \end{equation}
Consistency in the Kantarovich-Rubinstein metric in turn implies
consistency of the estimates of the distribution function at continuity points, which in turn implies consistency
of the estimates of the quantile function.

Equipped with the notations introduced above we can now state our result. \\

\begin{theorem}\label{theorem3}\textbf{(Consistent Estimation of Critical Values $k(1-\alpha)$).} Suppose
Conditions R.1-R.3 hold, and that we have a consistent estimate of the law of our limit Gaussian processes $V$.  Then the estimates of the  laws of the limit statistics $\mathcal{S}=\mathcal{W}$ or $\mathcal{L}$ defined above are consistent.  As a consequence, we have that the estimates of the quantiles are consistent in the sense of equation (\ref{quantile consistency}).
\end{theorem}

It is useful to give a similar result under more primitive conditions C.1-C.2. Recall that in this case our estimator satisfies
$$ \sqrt{n}(\widehat \gamma - \gamma) \rightsquigarrow \Omega^{1/2}Z,$$
so that our limit statistics take the form:
$$
\mathcal{S} = \sup_{\theta \in \partial \Theta_0} [V(\theta)]^2_+, \ \  V(\theta) = t(\theta)'Z,
$$
where $t(\theta)$ is a vector valued weight function, in particular, we have
$$
\begin{array}{ll}
t(\theta)= (\nabla_{\gamma} m(\theta, \gamma)'
\Omega^{1/2})/s(\theta) &  \text{ for } \mathcal{S}= \mathcal{L}, \\
t(\theta)= (\nabla_{\gamma}
m(\theta, \gamma)' \Omega^{1/2} )/ (\|
\nabla_{\theta} m(\theta, \gamma) \| \cdot
w(\theta))  & \text{ for } \mathcal{S}= \mathcal{W}.
\end{array}
$$
Here we shall
assume that we have a consistent estimate $\mathcal{Q}_{Z^*}$ of the law $\mathcal{Q}_Z$ of $Z$, in the sense that,
\begin{equation}\label{boot consistency gamma}
\rho_{K}( \mathcal{Q}_{Z^*}, \mathcal{Q}_Z) = o_p(1).
 \end{equation}
 For instance we can simulate the distribution using draws $Z^* \sim N(0,I)$ or apply any valid bootsrap method to $\widehat \Omega^{-1/2} \sqrt{n}(\widehat \gamma - \gamma)$. Alternatively it is possible to use subsampling (\cite{politis}). Then the estimate $\mathcal{Q}_{V^*}$ of the law  $\mathcal{Q}_{V^*}$ is defined as:
  \begin{equation}
 V^*(\theta) = \widehat t(\theta)'Z^*,
 \end{equation}
 where $\widehat t(\theta)$ is a vector valued weighting function that is uniformly consistent
 for the weighting function $t(\theta)$.   In particular, we can use
$$
\begin{array}{ll}
\widehat t(\theta)= (\nabla_{\gamma} m(\theta, \widehat \gamma)'
\widehat \Omega^{1/2})/\widehat s(\theta) &  \text{ for } \mathcal{S}= \mathcal{L}, \\
\widehat t(\theta)= (\nabla_{\gamma}
m(\theta, \widehat \gamma)' \widehat \Omega^{1/2} )/ (\|
\nabla_{\theta} m(\theta, \widehat \gamma) \| \cdot
w(\theta))  & \text{ for } \mathcal{S}= \mathcal{W}.
\end{array}
$$

 In this definition we hold the hatted quantity fixed, and the only random element being resampled or simulated is $Z^*$, with the law  denoted as $\mathcal{Q}_{Z^*}$.  Then, we define the
random variable
$$
\mathcal{S}^* = \sup_{\theta \in \widehat{ \partial \Theta_0}} [V^*(\theta)]^2_+,
$$
and use its law $\mathcal{Q}_{\mathcal{S}^*}$ to estimate the law $\mathcal{Q}_{\mathcal{S}}$.  In the definition of the law
$\mathcal{Q}_{\mathcal{S}^*}$ we fix, i.e. condition on, the estimated quantities $\widehat t(\cdot)$ and $\widehat{ \partial \Theta_0}$. \\

We can now state the following corollary, which is proven in the appendix:\\[8pt]

\begin{corollary}\label{corollary3} \textbf{(Consistent Estimation of Critical Values under Condition C).} Suppose that
conditions C holds, and that we have a consistent estimate of the law of $Z$, so equation (\ref{boot consistency gamma}) holds.   Then this provides us with a consistent estimate of the law of our limit Gaussian process $\Gp$, and hence all of the conclusions of Theorem 3 hold.\end{corollary}

\subsection{Verification of Condition C for MF- and HJ-Bounds}

We can now apply this result to the HJ and MF problems. As before, let $v:=\mathbb{E} [r_{t+1}]$ and $\Sigma:=\mathbb{E} [(r_{t+1}-v)(r_{t+1}-v)']$, and the parameters $\theta = (\mu,\sigma)'$, and $\gamma_{HJ}=(S_{vv},S_{v1},S_{11})'$ for the HJ bounds, where
\[S_{vv} = v'\Sigma^{-1}v,\;S_{v1} = 1_N'\Sigma^{-1}v,\;\textnormal{and }S_{11}=1_N'\Sigma^{-1}1_N.\]
Also let $\widehat{\gamma}_{HJ}$ denote the sample analog obtained by replacing $v$ and $\Sigma$ with the mean $\widehat{v}$ and variance $\widehat{\Sigma}$ of the sample $r_1,\dots,r_T$. For the MF bounds, let $\gamma_{MF}=\textnormal{vec}(D'\Sigma D)$, where
\[B:=\textnormal{Cov}_t(Z_{t+1},r_{t+1})\hspace{0.5cm}\textnormal{and }D:=[v,B']\]
and $\widehat{\gamma}_{MF}$ be the sample analog, and let $\mu^*$ and $\beta^*$ be the target mean portfolio return and factor loadings.

Recall that the Hansen-Jagannathan mean-variance bound for the stochastic discount factor is defined by the moment condition
\[m_{HJ}(\theta,\gamma_{HJ}) = \sqrt{S_{vv}\mu^2 - 2S_{v1}\mu + S_{11}}-\sigma.\]
whereas the multi-factor efficient (MF) mean-variance set for portfolio returns is characterized by
\[m_{MF}(\theta,\gamma_{MF}):=\sqrt{(\mu^*,\beta^{*'})(D'\Sigma D)^{-1}(\mu^*,\beta^{*'})'}-\sigma\]
We now give regularity conditions for the validity of inference and confidence intervals based on the LR and Wald statistics derived from the moment functions $\widehat{m}_{HJ}(\theta)$ and $\widehat{m}_{MF}(\theta)$, respectively.

\begin{proposition}[\textbf{Verification of Condition C for HJ and MF problems}]\label{hj_m_c1_2_prp}
Suppose that $\widehat{v},\widehat{\Sigma}$, and $\hat{B}$ satisfy a CLT. Furthermore, let $\Theta$ be a rectangle in $\mathbb{R}\times\mathbb{R}_+$, that the absolute values of the elements of $v$ and eigenvalues of $\Sigma$ are bounded between finite strictly positive constants, and that  $N\ssum_{i=1}^N v_i^2 - \left(\ssum_{i=1}^Nv_i\right)^2$ is bounded away from zero. Then Condition C holds for the HJ and MF bounds.
\end{proposition}

For the last condition, note that for the purposes of this paper, we treat the number of assets $N$ as finite, and that the Cauchy-Schwarz inequality implies that the difference $N\ssum_{i=1}^N v_i^2 - \left(\ssum_{i=1}^Nv_i\right)^2\geq0$. The difference will be strictly positive only if the mean return $v_i$ is not constant across assets.

\subsection{Structured Projection Approach}

We also compare our procedure to alternative confidence regions from a structured projection approach that is based on a confidence set for the point-identified parameter $\gamma$ that characterizes the bound $m(\theta) = m(\theta, \gamma)$ under the condition C.
This confidence region is then ``projected" to obtain a confidence region for $\Theta_0$.  We call the projection approach ``structured" when $\gamma$ is a minimal sufficient parameter for the bounds on $\theta$, and
$\widehat \gamma$ is the minimal sufficient statistics. For example the HJ bounds from Example 1 can be characterized by the three-dimensional parameter $$\gamma = (S_{vv}, S_{v1}, S_{11})',$$ whereas the Markowitz-Fama mean-variance frontiers are described in terms of the first two moments of the spanning tangency portfolios.   Note that the structured approach avoids creating some confidence regions for high-dimensional
mean and variance parameters $v$ and $\Sigma$ and then projecting them to obtain a confidence region for $\Theta_0$. Such an approach would be extremely conservative, and working with the minimal parameter $\gamma$ instead reduces the conservativeness dramatically.
Since inference for $\gamma$ is standard, the structured projection approach seems much more natural and ``economically appealing". However, despite its dimension-reducing and intuitive appeal, this approach remains very conservative and is much less powerful than
the approach based on the optimally weighted LR-type and W-statistics.  We illustrate the superior performance of LR-based confidence sets compared to the structured projection approach empirically for the Hansen-Jagannathan set (see Figure~\ref{projection_graphs}).

For the construction of projection confidence sets, we assume that we can construct a $(1-\alpha)$ confidence set $R_{\gamma}$ for $\gamma_0$. This confidence region can then be projected onto the parameter space $\Theta$ to form the confidence set for $\Theta_0$:
\begin{equation}\label{R-proj}
R_{Proj}:=\bigcup_{\gamma\in R_{\gamma}} \left\{\theta\in\Theta: m(\theta,\gamma)\leq 0\right\}.
 \end{equation}
Under condition C, $\sqrt{n}(\widehat{\gamma}-\gamma_0)$ satisfies a CLT, so we can construct an elliptical joint $1-\alpha$ confidence region for the quantity $\gamma$ as follows:
\begin{equation}\label{R-gamma}
R_{\gamma} = \{ \gamma \in \Gamma: n(\widehat{\gamma}-\gamma)'\widehat \Omega^{-1}(\widehat{\gamma}-\gamma) \leq \widehat c(1-\alpha)  \},
\end{equation}
where $\widehat c(1-\alpha)$ is either $1-\alpha$-quantile of $\chi^2(k)$ variable, or any consistent estimate of such a quantile.
Under asymptotic normality of $\widehat{\gamma}$, this construction approximates an upper contour set of the density of the estimator, and therefore gives an (approximate) smallest-volume confidence set for the parameter $\gamma$.

\begin{proposition}\label{projection_approach_val}\textbf{(Basic Validity of Structured Projection Approach).} Let $R_{\gamma}$ be  $1-\alpha$ confidence set for the parameter $\gamma$. Then $\Pp[\Theta_0\subset R_{proj}]\geq \Pp[\gamma_0\in R_{\gamma}] = 1-\alpha$.
In particular, under Condition C, the region $R_\gamma$ in (\ref{R-gamma}) obeys
$$
\Pp[\gamma_0 \in \Gamma] = 1-\alpha + o(1),
$$
so that the confidence region (\ref{R-proj}) based on structured projection obeys
$$
\Pp[\Theta_0\subset R_{proj}] \geq 1- \alpha + o(1).
$$

\end{proposition}

Projected confidence sets were first proposed by \cite{Sch1953} for the problem of joint confidence bounds for all linear combinations of the form $c'\gamma$, where $\gamma\in\mathbb{R}^d$ is a parameter vector and $c\in\mathbb{S}^{d-1}$, the $d-1$ dimensional unit sphere in $\mathbb{R}^d$. For the problem of confidence bands for the linear regression function, it has been shown that optimality of \cite{Sch1953}'s method depends crucially on equivariance with respect to translations and orthogonal transformations of the original parameter, see sections 9.4 and 9.5 in \cite{LRo2005} for a discussion. \cite{Boh1973} showed optimality of projection bounds for a generalization of \cite{Sch1953}'s original problem, but they were shown to be suboptimal if coverage was only required for a restricted set of functionals corresponding to vectors $c$ for a proper subset of $\mathbb{S}^{d-1}$, see \cite{CSt1980} and \cite{Nai1984}.

In our case, there are at least three regards in which equivariance with respect to the reduced-form parameters $\gamma$ fails: in all our examples, (1) the dimension of $\theta$ is strictly lower than that of $\gamma$, (2) we consider inference problems that are one-sided rather than symmetric where the parameter space is only a compact subset of $\mathbb{R}^k$, and (3) the inequality generating function $m(\theta,\gamma)$  is nonlinear in $\gamma$.  Even after approximate linearization, our inference problem reduces to inference on $c'\gamma$ with $c$ restricted to a small subset of the sphere  $\mathbb{S}^{d-1}$, which is exactly the case where the structured projection approach becomes suboptimal. As mentioned above, we illustrate the superior performance of LR-based confidence set compared to the structured projection based set empirically for the HJ problem (see Figure~\ref{projection_graphs}) -- the LR-based set is much smaller and lies strictly inside the projection-based set.

\subsection{Invariance and Similarity Properties of Confidence Regions based on LR and Wald Statistics}

We next proceed to state the invariance properties of the proposed inference procedures with respect to parameter transformations. We distinguish between an exact invariance and an asymptotic invariance. A parameter transformation is a one-to-one mapping $\eta:\Theta\rightarrow\Upsilon$,
 where $\Upsilon = \eta(\Theta)$, and we denote the population and sample moment conditions for the transformed problem by $m_{\eta}(\cdot):=m(\eta^{-1}(\cdot))$,   and $\widehat{m}_{\eta}(\cdot):=\widehat{m}(\eta^{-1}(\cdot))$, respectively,
 which are mappings from $\Upsilon \to \Bbb{R}$.    We define $H^*$ as the set of parameter transformations $\eta$ that are continuously
  differentiable in $\theta \in \Theta^{\delta}$ and such that Conditions R.1-R.3 hold for the transformed moments $m_{\eta}(\cdot)$.

In this section we discuss invariance properties of inference based on the LR and W statistics. Let \[\mathcal{L}_n(\theta): = \left[\sqrt{n}\widehat{m}_n(\theta)/\widehat{s}(\theta)\right]_+^2\hspace{0.5cm}\textnormal{ and  }\mathcal{W}_n(\theta): = \left(\sqrt{n}d(\theta,\widehat{\Theta}_{0,\delta})/\widehat{w}(\theta)\right)^2,\] and let the critical values $\widehat{k}_L(1-\alpha)$ and $\widehat{k}_W(1-\alpha)$  be consistent estimators of the asymptotic $1-\alpha$ quantiles of $\mathcal{L}_n=\sup_{\theta\in\Theta_0}\mathcal{L}_n(\theta)$ and $\mathcal{W}_n=\sup_{\theta\in\Theta_0}\mathcal{W}_n(\theta)$, respectively. Consider the decision functions for including a value of $\theta$ in the confidence regions based on the LR and Wald-type statistics, respectively, \[\phi_{Ln}(\theta):=\mathbf{1}\{\mathcal{L}_n(\theta)\leq \widehat{k}_L(1-\alpha)\}\ \textnormal{ and  } \
\phi_{Wn}(\theta):=\mathbf{1}\{\mathcal{W}_n(\theta)\leq \widehat{k}_W(1-\alpha)\},\] respectively. Similarly, given the parameter transformation $\eta$, we consider the statistics
\[\mathcal{L}_n(\theta;\eta): = \left[\sqrt{n}\widehat{m}_{\eta }(\eta(\theta))/\widehat{s}(\theta)\right]_+^2\ \textnormal{ and  }\ \mathcal{W}_n(\theta;\eta): = \left(\sqrt{n}d(\eta(\theta),\eta(\widehat{\Theta}_{0,\delta}))/\widehat{w}(\theta;\eta)\right)^2\]
with a weight function $\widehat{w}(\theta;\eta)$ possibly depending on $\eta$, and the resulting decision functions
\[\phi_{Ln}(\theta;\eta):=\mathbf{1}\{\mathcal{L}_n(\theta;\eta)\leq \widehat{k}_{L\eta}(1-\alpha)\}\ \textnormal{ and  } \
\phi_{Wn}(\theta;\eta):=\mathbf{1}\{\mathcal{W}_n(\theta;\eta)\leq \widehat{k}_{W\eta}(1-\alpha)\},\] where the critical values $\widehat{k}_{L\eta}(1-\alpha)$ and $\widehat{k}_{W\eta}(1-\alpha)$ are estimates of the asymptotic $1-\alpha$ quantiles of $\mathcal{L}_n(\eta)=\sup_{\theta\in\partial\Theta_0}\mathcal{L}_n(\theta;\eta)$ and $\mathcal{W}_n(\eta)=\sup_{\theta\in\partial\Theta_0}\mathcal{W}_n(\theta;\eta)$, respectively.

We say that the decision function $\phi_n(\theta;\eta)$ is
\begin{itemize}
\item \emph{invariant} if for any $\theta\in\Theta$, we have $\phi_n(\theta)=\phi_n(\theta;\eta)$ for any parameter transformation $\eta\in H^*$,
\item \emph{asymptotically invariant} to first order if for any parameter transformation $\eta\in H^*$, any $\theta_0\in\Theta$, and any sequence $\theta_n\rightarrow\theta$, we have $P\left(\phi_n(\theta_n)\neq\phi_n(\theta_n;\eta)\right)\rightarrow0$ as $n\rightarrow\infty$.
\end{itemize}
These invariance properties describe whether the parameter transformations affect the inclusion of any sequence of points $\theta_n$ in the confidence sets. Note in particular that (exact) invariance is a property of a given realization of $\phi_n(\theta;\eta)$ and implies asymptotic invariance. Furthermore, it is easy to verify that invariance of a decision function $\phi_n(\theta;\eta)$ implies analogous equivariance properties for the corresponding confidence sets \[R_{LR}\equiv\{\theta\in\Theta:\phi_{Ln}(\theta;\eta)=1\}\hspace{0.5cm}\textnormal{ and }R_{W}\equiv\{\theta\in\Theta:\phi_{Wn}(\theta;\eta)=1\}.\]

We also would like to mention another property, which characterizes the precision of the confidence sets.  For a given value of $\eta$, we also say that a test based on $\phi_n(\theta;\eta)$ is \emph{asymptotically similar} on the boundary of $\Theta_0$ if $\lim_n\mathrm{E}[\phi_n(\theta;\eta)]=c$ for some constant $c \in (0,1)$ and any  $\theta \in \partial \Theta_0$.  Similarity here means that any point on the boundary of the identified set can be expected to be included with asymptotic probability $c$, which does not vary with the location of the point.

Given these definitions, we can now characterize the invariance and similarity properties of inference procedures based on the weighted LR and Wald statistics:

\begin{proposition}\label{invariance_prop}\textbf{(Invariance Properties of Decision Functions based on $\mathcal{L}_n(\theta;\eta)$ and $\mathcal{W}_n(\theta;\eta)$).} Suppose Conditions R.1-R.3 hold. Then (i) the decision function $\phi_{Ln}(\theta;\eta)$ based on $\mathcal{L}_n(\theta;\eta)$ is invariant, and asymptotically similar on the boundary of $\Theta_0$, whereas (ii) the decision function $\phi_{Wn}(\theta;\eta)$ based on $\mathcal{W}_n(\theta;\eta)$ is asymptotically invariant to first order if the weighting function is of the form $\widehat{w}(\theta;\eta)=\frac{b(\theta)}{\|\nabla_{\eta}m_{\eta}(\eta(\theta))\|}+o_p(1)$ for some function $b(\theta)>0$ that does not depend on $\eta$ and all $\theta\in\partial\Theta$. Furthermore, (iii) if $\widehat{w}(\theta;\eta)=b\frac{\widehat{s}(\theta)}{\|\nabla_{\eta}m_{\eta}(\eta(\theta))\|} + o_p(1)$ for some constant $b>0$, then $\phi_{Wn}(\theta)$ is asymptotically similar on the boundary of $\Theta_0$.
\end{proposition}

Notice in particular the different roles the norm of the gradient of $m(\theta)$ and the standard deviation $s(\theta)$ play for the properties of the weighted Wald statistic: Choosing weights that are inversely proportional to $\|\nabla_{\theta}m(\theta)\|$ in the limit corrects for the dependence of the Hausdorff distance on the parameterization of the problem, and accounting for $s(\theta)$ also gives similarity on the boundary. In particular, only the weights in part (iii) of Proposition \ref{invariance_prop} yield results for confidence sets based on the Wald statistic that compare to the performance of LR-based inference. Our empirical results below illustrate that the difference is important, since the lack of invariance or precision can lead to overturning the main economic conclusions in the empirical analysis.

\section{Empirical Applications}

\subsection{Hansen-Jagannathan Mean-Variance Sets for the SDF}
In order to illustrate the performance of our procedure, we estimate confidence sets for the Hansen-Jagannathan sets of mean-variances of stochastic discount factors. In order to keep results comparable with \cite{HJ1991}, we construct the sample for the empirical exercise following the data description in \cite{HJ1991}. The two asset series used are annual treasury bond returns and annual NYSE value-weighted dividend included returns. These nominal returns are converted to real returns by using the implicit price deflator based on personal consumption expenditures used by \cite{HJ1991}. Asset returns are from CRSP, and the implicit price deflator is available from St. Louis Fed and based on National Income and Product Accounts of United States. We use data for the years 1959-2006.

\begin{figure}[p]
\centering
\includegraphics[width=100mm, height=100mm]{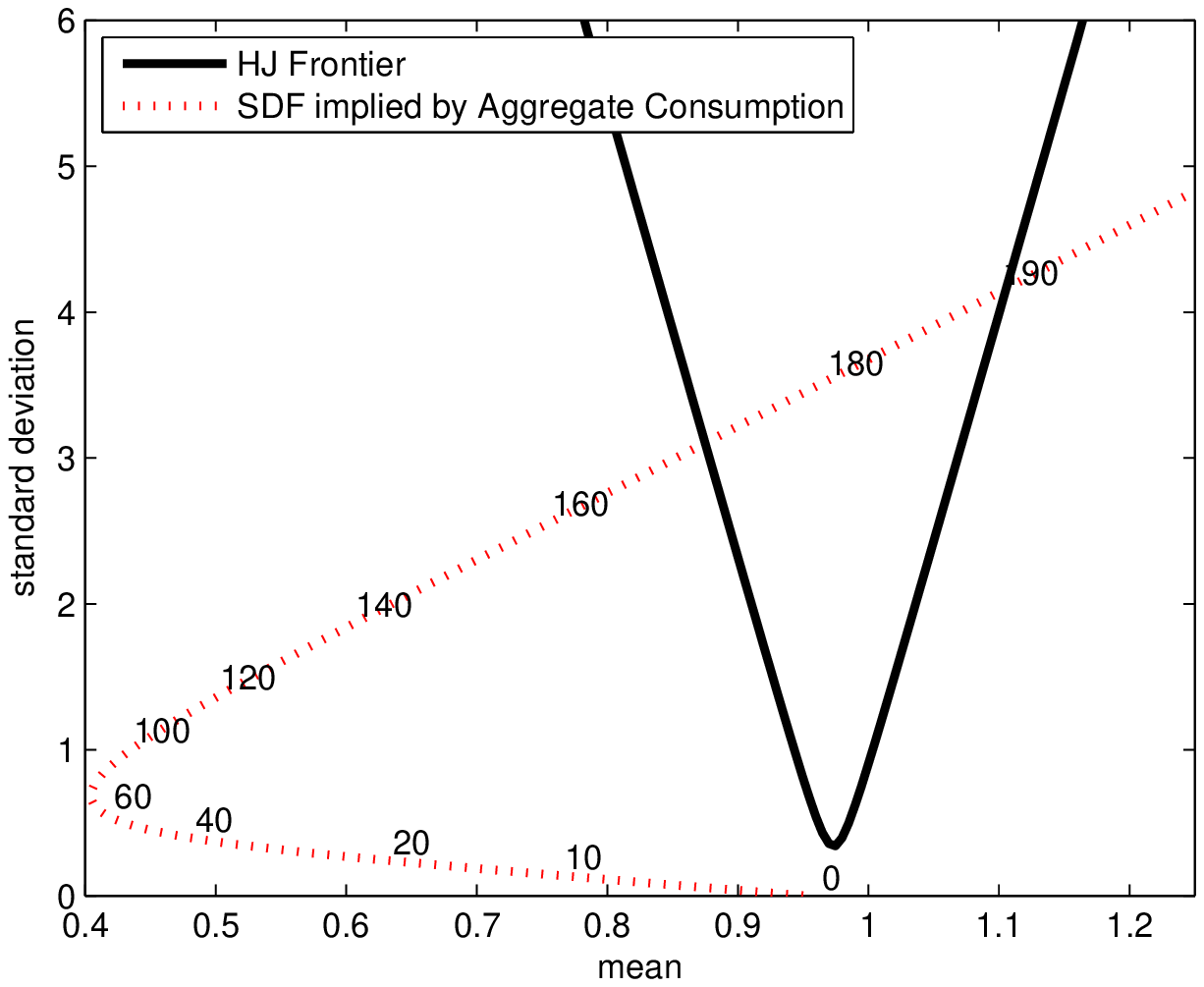}
\caption{Estimated HJ Bounds with Stochastic Discount Factors implied by CRRA preferences, data labels represent values for the IES $\varrho$}\label{alonefrontier}
\includegraphics[width=100mm, height=100mm]{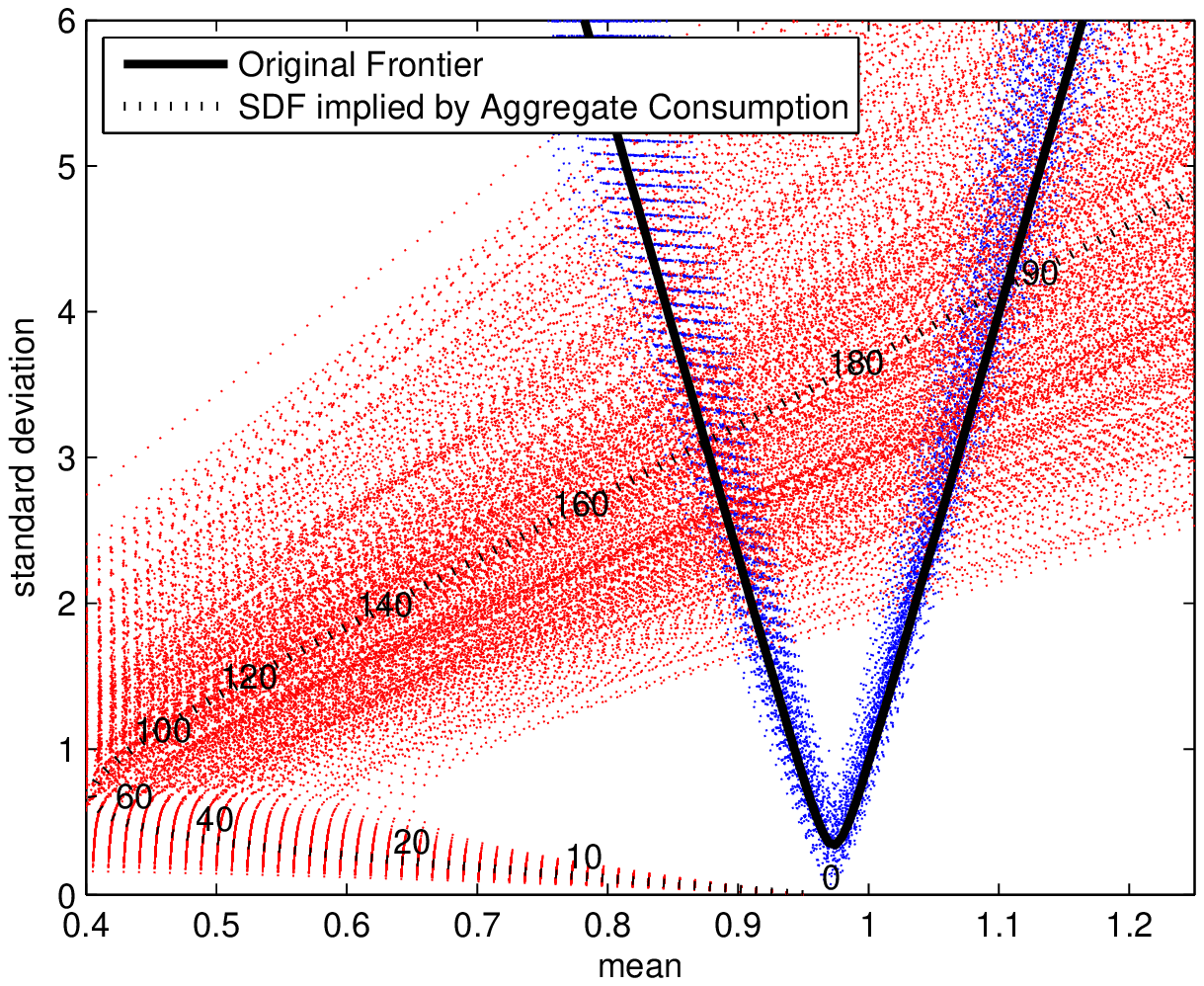}
\caption{Estimated HJ Bounds, Mean-Standard deviation pairs based on Consumption SDF, and Bootstrap Draws}\label{100BootstrappedFrontiers}
\end{figure}

Figure~\ref{alonefrontier} reports the estimated bound consisting of the mean-standard deviation pairs which satisfy
\begin{equation*}
m\left( \theta ,\widehat{\gamma}\right) =0,
\end{equation*}
where $\widehat{\gamma}$ is estimated using sample moments.

We can compare the estimated HJ bounds with mean-variance combinations implied by the consumption CAPM model. In the model the economy is equivalent to a representative agent with constant elasticity of intertemporal substitution preferences
\[u(C_t) =\frac{C_t^{1-\varrho}}{1-\varrho},\]
where $C_t$ is the aggregate consumption.  Then the stochastic discount factor implied by consumption growth is given by $M_t(\varrho) = \beta\left(\frac{C_t}{C_{t+1}}\right)^\varrho$, so that we can estimate the first two moments of $M_t(\varrho)$. Specifically, let $\mu_{C}(\varrho):=\mathbb{E}\left[M_t(\varrho)\right]$ and $\sigma_{C}(\varrho):=\sqrt{\textnormal{Var}\left(M_t(\varrho)\right)}$ be the mean and standard deviation, respectively, of the stochastic discount factor given an iso-elastic utility function with an elasticity of intertemporal substitution equal to $1/\varrho$. We can characterize the feasible set $\Theta_e$ of mean-variance pairs by the moment restriction
\[0 = m((\mu,\sigma)): = \mu_{C}(\sigma_{C}^{-1}(\sigma)) -\mu.\]
where $\sigma_C^{-1}(\cdot)$ denotes the inverse function of $\sigma_C(\varrho)$.\footnote{It can be verified that $\sigma_C(\varrho)$ is strictly increasing in $\varrho$, so that this inverse is well-defined.} Given an i.i.d. sample of observations for the growth rate of consumption $\frac{C_{t+1}}{C_t}$, we define the empirical analog $\widehat{m}((\mu,\sigma))$ analogously, where the expectations in the definition of $\mu_C(\cdot)$ and $\sigma_C(\cdot)$ are replace with averages. The mean-variance pairs reported in Figure~\ref{alonefrontier} were obtained using data on per capita expenditures for non-durable consumption and services in the U.S. from 1959-2006, assuming a discount factor $\beta=0.95$.

It is well known that it is difficult to reconcile asset prices and aggregate consumption empirically in a representative framework. For our data, the values of $\varrho$ corresponding to mean-variance pairs for the SDF that fall inside the estimated HJ set range from about 170 to 192, suggesting a very low elasticity of intertemporal substitution and implying a high variance for the SDF. These values for $\varrho$ are unrealistically large, but are in line with other findings in the empirical literature on the consumption based CAPM.\footnote{E.g. in \cite{LL2009}, the values for the elasticity of intertemporal substitution minimizing the mean squared error in the Euler equations characterizing the household's investment problem are comparable in magnitude.}

In order to represent the sampling uncertainty in estimating $\gamma$, we plot 100 bootstrap draws of the HJ frontier in Figure~\ref{100BootstrappedFrontiers}, where observations were drawn with replacement from the bivariate time series of stock and bond returns. In order to represent the sampling uncertainty in estimating the mean-standard deviation pairs $(\mu_C(\varrho),\sigma_C(\varrho))$
of the consumption-based SDF, we also plot 100 bootstrap draws of $\varrho \mapsto (\widehat \mu_C(\varrho), \widehat \sigma_C(\varrho))$.  We see that the sampling uncertainty is quite considerable for both HJ frontier and for mean-standard deviation pairs implied by the consumption-based SDF. In fact, an intriguing feature of this graph is that near the apex of the HJ frontier, the pair $(\widehat \mu_C(\varrho),\widehat \sigma_C(\varrho))$ and its bootsrap draws are close to the HJ frontier at low values of $\varrho$. The low values of $\varrho$ are considered to be  "reasonable," since they correspond to a relatively high intertemporal elasticity of substitution and appear to be well micro-founded . The inability of ``reasonable" values of $\varrho$ to reconcile aggregate consumption data with asset prices has been a major theme of the empirical literature on the consumption-based CAPM model starting with \cite{HS1982}. However from Figures 1 and 2 it is not obvious that an empirical test will reject the canonical/baseline model underlying that literature, and we report confidence sets based on the various approaches discussed in earlier sections based on which we can make inferential statements about the benchmark model.

\begin{figure}[p]
\centering
\includegraphics[width=100mm, height=100mm]{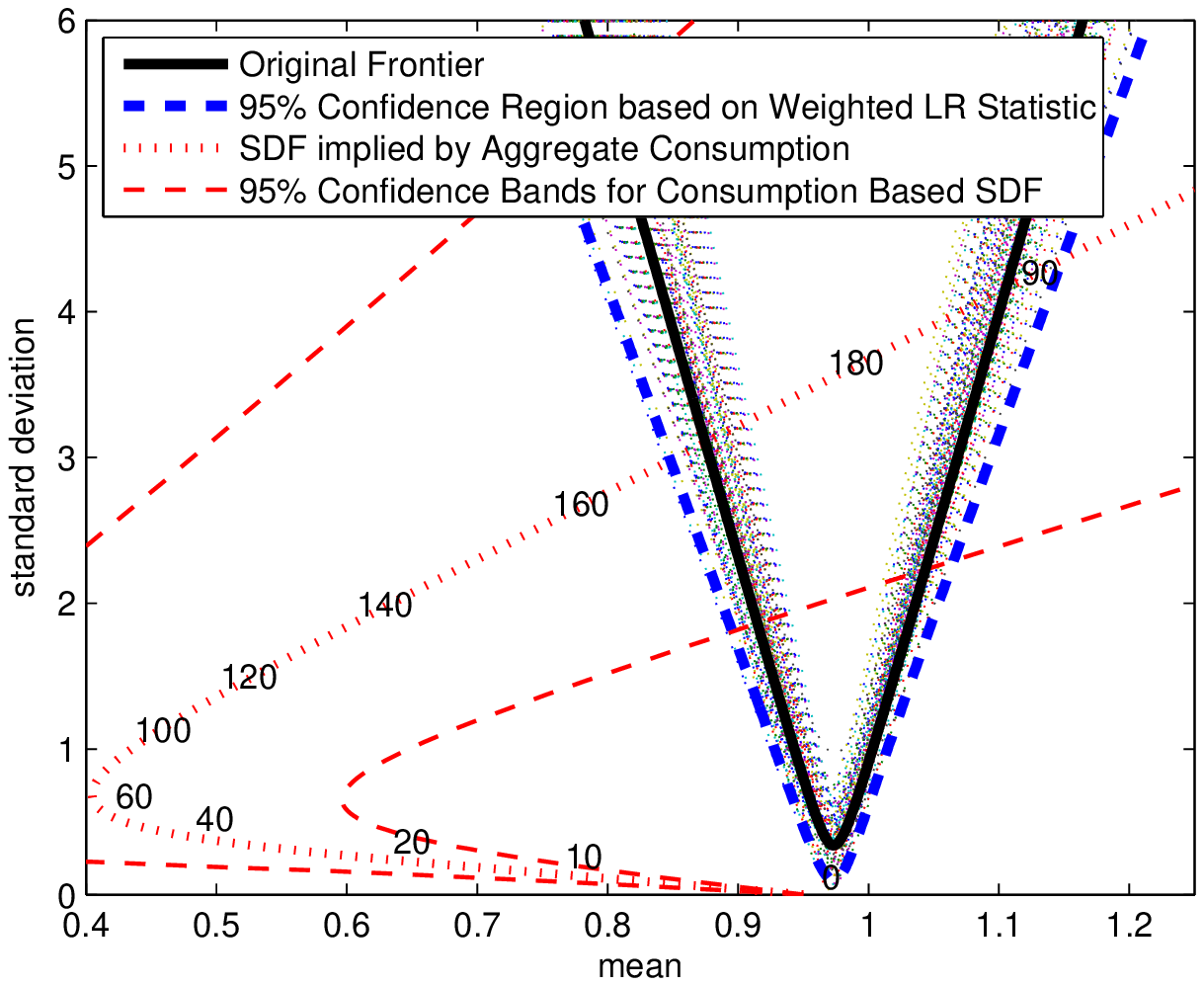}
\caption{95\% Confidence Set for the HJ Set using LR Statistic. The 95\% confidence set for mean and standard deviation of the consumption-based SDF was also constructed using the LR Statistic for the moment equality defining the corresponding mean-variance pairs.}
\label{weightedverticaldistancewithbootstrappedcurves}
\includegraphics[width=100mm, height=100mm]{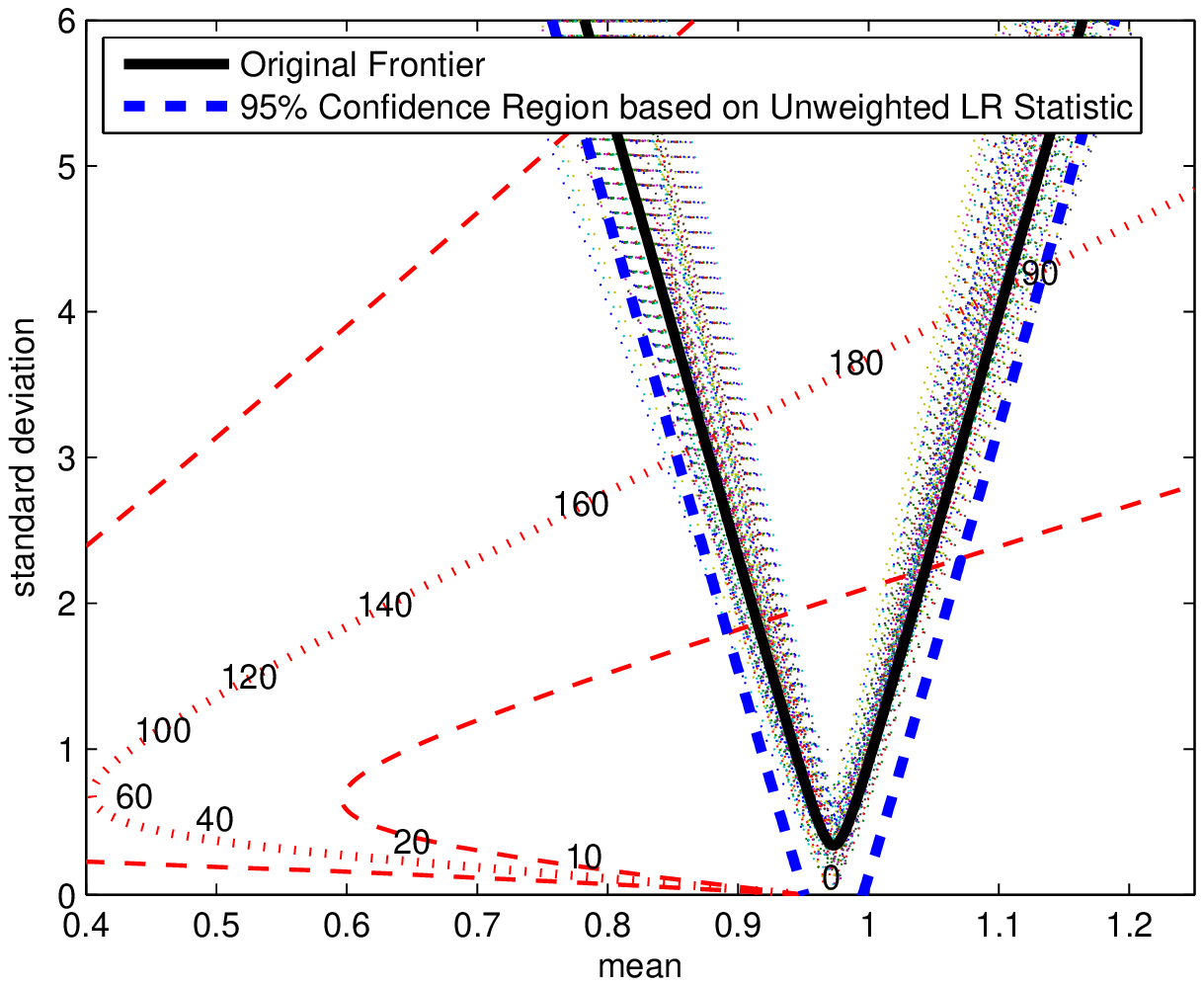}
\caption{95\% Confidence Set for the HJ Set using Unweighted LR Statistic}
\label{unweightedverticaldistancewithbootstrappedcurves}
\end{figure}

Figure~\ref{weightedverticaldistancewithbootstrappedcurves} shows the 95\% confidence region based on the LR statistic. By construction, the LR confidence region covers most of the bootstrap draws below the HJ bounds. However, it should also be noted that the confidence bound based on the LR statistic is fairly tight relative to the boostrapped frontiers, and does not include any unnecessary areas of the parameter space. Noting that the set $\Theta_e$ for the consumption-based SDF is defined by a moment equality, we can also form a confidence band $R_e$ for $\Theta_e$ based on the statistic $\mathcal{L}_n(\theta):=\left[\frac{\widehat{m}_{SDF}(\theta)}{\widehat{s}(\theta)}\right]^2$, noting that the asymptotic arguments in the derivations for the (one-sided) LR statistic can be easily extended to the two-sided case if we replace squared positive parts $[\cdot]_+^2$ with the usual square, $(\cdot)^2$. The lower and upper bounds in the following figures were constructed using separate estimates of the local standard deviation $s((\mu,\sigma))$ based on the negative and positive deviations of $\widehat{m}((\mu,\sigma))$, respectively to improve the approximation. Critical values were obtained using the nonparametric bootstrap.

Most importantly, the LR-based confidence region for the HJ set does not overlap with the confidence set for the consumption-based SDF for ``small" values of $\varrho$ (in fact, for any $\varrho\in[0,120]$). The absence of overlap for the $95\%$ regions implies the rejection
of any  $\varrho\in[0,120]$ at $10\%$ significance level. This is clear evidence against the benchmark formulation of the consumption-based CAPM, and therefore an important empirical conclusion.  In what follows below we will show that the same empirical conclusion
cannot be reached for this example using less precise or non-invariant methods. Specifically, we will show that if
we use confidence regions based on either LR-statistic without precision weighting, or Wald statistics without invariance/precision
weighting, or regions based on structural projection, we will not be able to reach the same empirical conclusion.  So invariance
and precision considerations in construction of the confidence regions turn out to be quite important for reaching sharp economic conclusions.

Figure~\ref{unweightedverticaldistancewithbootstrappedcurves} plots the 95\% confidence region
based on an unweighted LR statistic. Comparing Figure~\ref{weightedverticaldistancewithbootstrappedcurves} and
Figure~\ref{unweightedverticaldistancewithbootstrappedcurves} it can be seen that precision weighting
plays a very important role in delivering good confidence sets. Without precision weighting, the
unweighted LR statistic delivers a confidence region that includes implausible regions in the parameter space where the
standard deviation of the discount factor is zero. Moreover, the confidence region becomes too imprecise
to reject the canonical model of the stochastic discount factor.

\begin{figure}[p]
\centering
\includegraphics[width=100mm, height=100mm]{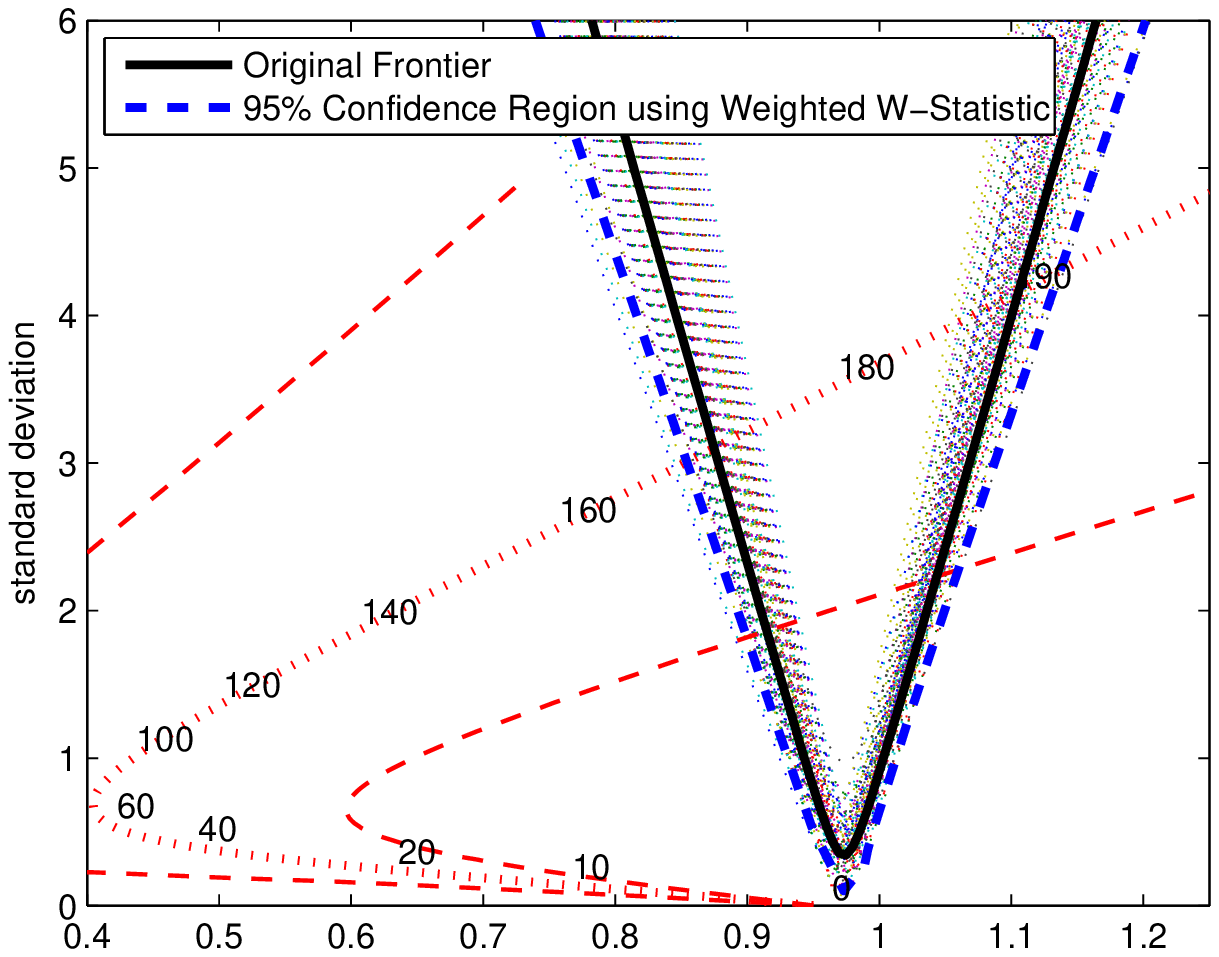}
\caption{95\% Confidence Set for the HJ Set using Weighted W Statistic}
\label{weightedhausdorffdistancewithbootstrappedcurves}
\includegraphics[width=100mm, height=100mm]{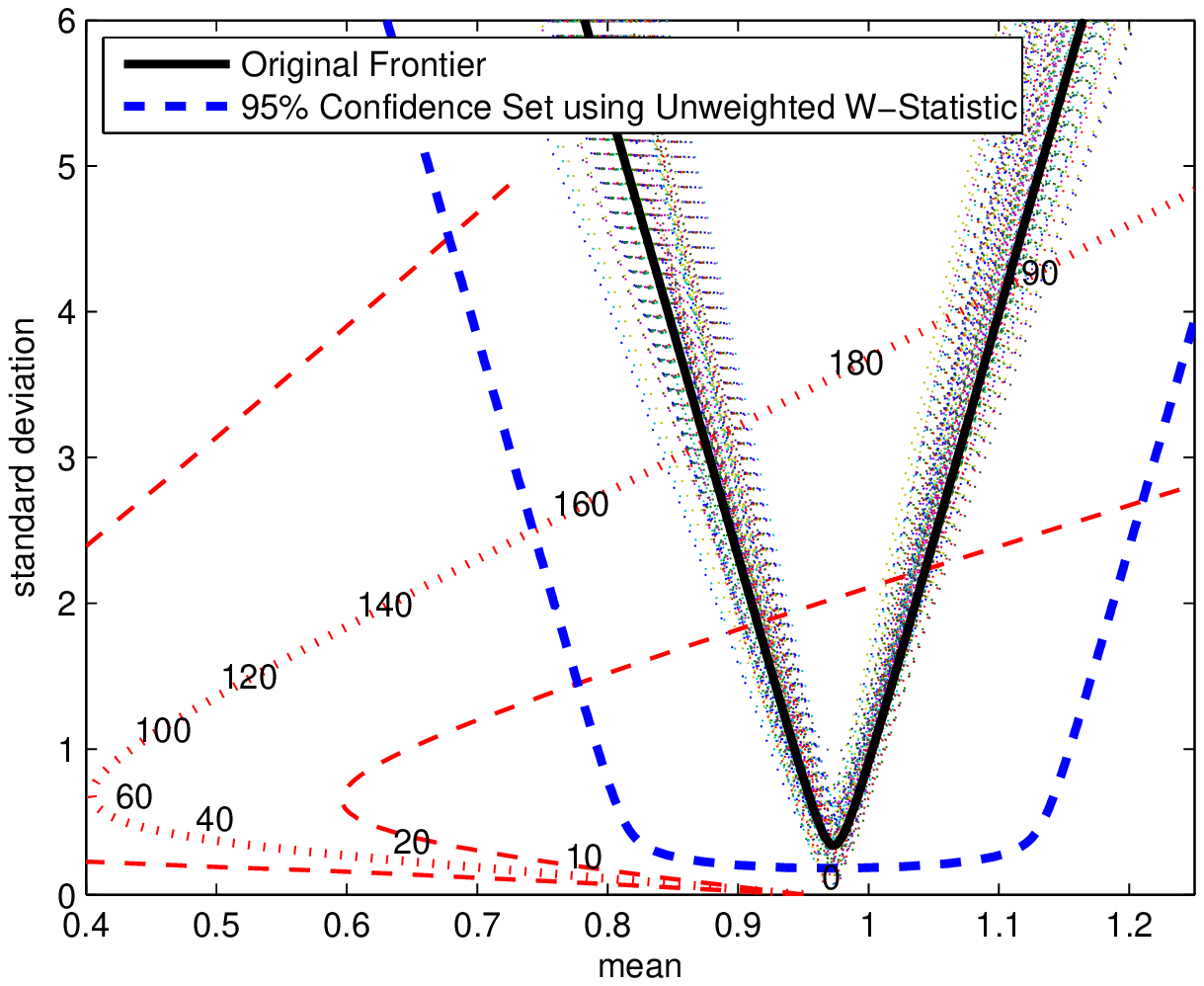}
\caption{95\% Confidence Set for the HJ Set using Unweighted W Statistic (directed Hausdorff Distance)}
\label{hausdorffdistancewithbootstrappedcurves}
\end{figure}

\begin{figure}[p]
\centering
\includegraphics[width=100mm, height=100mm]{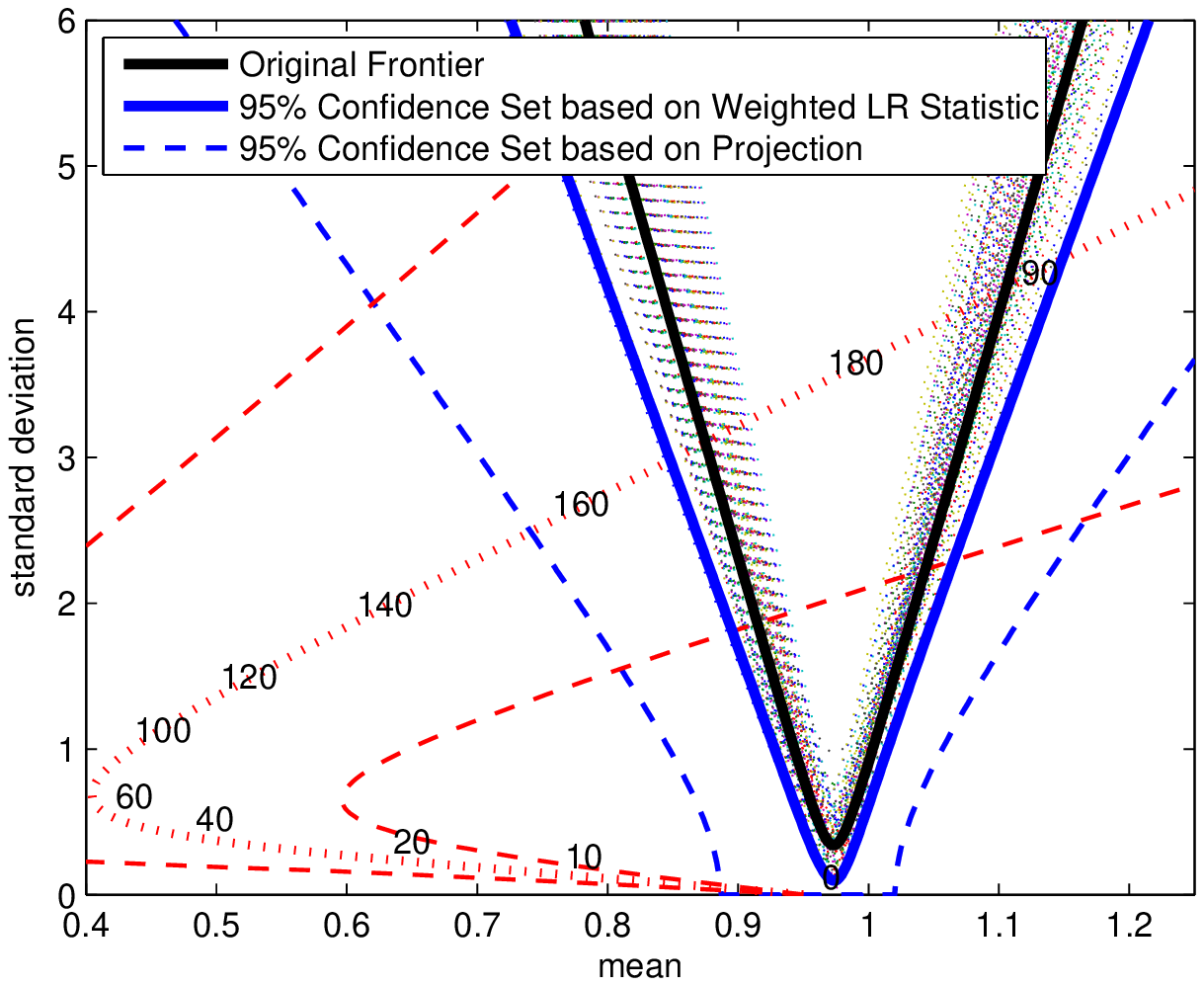}
\caption{95\% Confidence Sets for the HJ Set using Weighted LR Statistic and Projection Approach}
\label{projection_graphs}
\includegraphics[width=100mm, height=100mm]{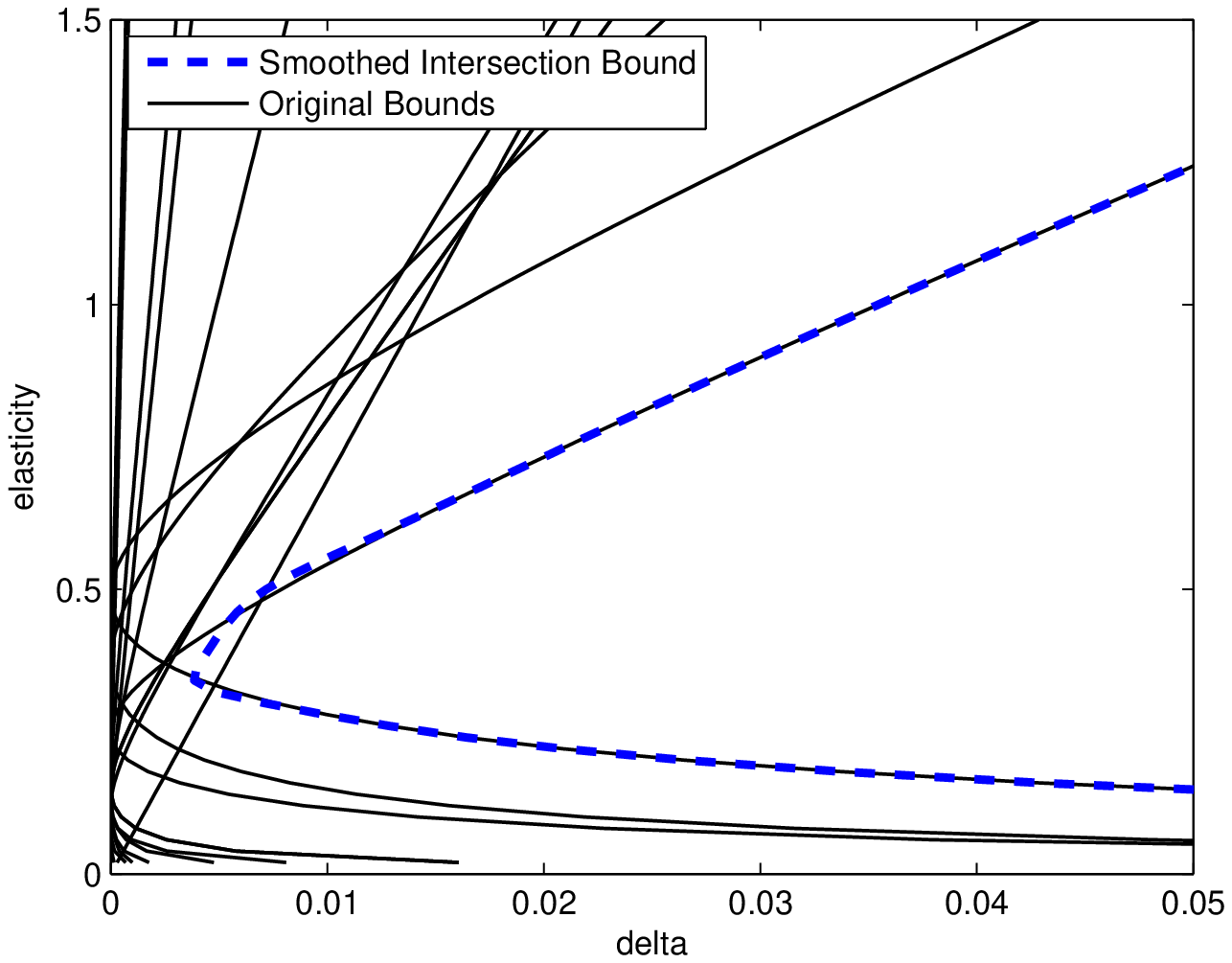}
\caption{Bounds on Structural Labor Supply Elasticity based on Estimates from Different Studies, and their Smoothed Intersection Bound}\label{elast_bound}
\end{figure}

Figure~\ref{hausdorffdistancewithbootstrappedcurves} plots the confidence region based on the Wald statistic with no invariance/precision weighting, which is equivalent to a confidence region based on the directed Hausdorff distance. Similar to Figure~\ref{unweightedverticaldistancewithbootstrappedcurves} the confidence set covers a large area of the parameter space which is excluded from any bootstrap realization of the HJ set. The shape of the confidence region based on the Wald statistic in Figure~\ref{hausdorffdistancewithbootstrappedcurves} seem counter-intuitive because at first sight, as the confidence bounds do not appear to be a uniform enlargement of the estimated frontier $\partial\widehat{\Theta}$. However, this visual impression is only due to the fact that the plot shows units of $\mu$ and $\sigma$ at different scales. The observation that the weighting and scaling of the different components of $\theta$ seem ``unnatural" in this particular graph emphasizes the potential problems associated with the non-invariance of inference based on the unweighted Wald statistic.

Figure~\ref{weightedhausdorffdistancewithbootstrappedcurves} plots the confidence region based on the
weighted Wald statistic, where weights induce first order invariance and similarity via precision weighting. This weighting fixes the problem and generates a statistic that is (first-order) invariant to parameter transformations. As a result, the  confidence set looks very similar to
weighted LR based confidence set in Figure~\ref{weightedverticaldistancewithbootstrappedcurves} in that
it covers most of the bootstrap draws below the HJ bounds and its shape reflects local sampling uncertainty in an adequate manner.
This practical evidence therefore emphasizes the importance of introducing invariance and precision inducing weights in the Wald-based approach, which we had argued for theoretically in the previous sections.

Finally, in Figure 7 we compare our results to confidence regions from the structured projection approach that is based on a confidence set for the point-identified parameters. As described in section 2.5, we construct an elliptical joint $1-\alpha$ confidence region for the quantity $\gamma$ defined in equation (\ref{define ABC}) based on the quadratic form for the estimator $\widehat{\gamma}$,  $\widehat{T}(\gamma):=(\widehat{\gamma}-\gamma)'\textnormal{var}(\widehat{\gamma})^{-1}(\widehat{\gamma}-\gamma)$. For the diameter of this confidence ellipsoid we used both a bootstrap and a chi-square approximation to the distribution of $\widehat{T}(\gamma)$, which both yield qualitatively similar results.

The $1-\alpha$ confidence set for $\theta=(\mu,\sigma)$ is obtained by projecting the $1-\alpha$ confidence region for $\gamma$ onto $\Theta$ using the condition $m(\theta,\gamma)\leq0$. We report the resulting confidence set from the structured projection approach in Figure~\ref{projection_graphs} together with the LR-based confidence set proposed in this paper. The structured projection confidence set performs quite poorly relative to the LR-based confidence set: in particular the latter is much smaller and lies strictly inside the former. In fact, the precision of the confidence set based on structured projection is poor enough to overturn the major empirical conclusion that the consumption-based CAPM cannot be reconciled with small values of $\varrho$.

This should be expected since the projection confidence bounds are based on a confidence set for the point-identified parameter that does not account for the specific shape of the bounds as a function of $\widehat{\gamma}$. More specifically, the elliptical joint confidence set for $\gamma$ (which minimizes volume under joint normality of $\widehat{\gamma}$) guards us against deviations from the true value in any direction in $\mathbb{R}^3$, but most of these deviations are irrelevant for the bounds for $\theta$, since these are only one-sided and the parameter space for $\theta$ is only two-dimensional. The fact that the standard confidence set for $\gamma$ treats all directions in the parameter space $\Gamma$ symmetrically may be far from ideal for inference on the $(\mu,\sigma)$-frontier, since the bound on the standard deviation is a nonlinear function whose derivative with respect to $\gamma$ varies widely across different values of $(\mu,\sigma(\mu))$. Note that for confidence sets for a point-identified parameter,  by the delta method the effect of nonlinearities is asymptotically negligible to first order. However when the object of interest is a set with a nontrivial diameter, the resulting effect is of first order even for large samples.


\subsection{Bounds on the Elasticity of Labor Supply}

In his meta-analysis, \cite{C2012} reports point-wise confidence bounds for the structural Hicksian elasticity $\varepsilon$ of labor supply at the intensive margin for given values of the optimization friction $\delta$. The reported bounds result from the intersection of bounds of the form (\ref{chetty_bound}) from estimates $\widehat{\varepsilon}_j$ obtained from $J$ empirical studies studies exploiting different natural experiments varying the effective income tax $\Delta_j\log p$, $j=1,\dots,J$.

We apply the bootstrap procedure proposed in this paper to obtain joint confidence sets for $(\delta,\varepsilon)$ based on the LR and Wald statistics. More specifically, we consider the moments obtained from individual empirical elasticities $\widehat{\varepsilon}_j$
\[\widehat{g}_{OF,j}((\varepsilon,\delta)')=g_{OF,j}((\varepsilon,\delta)',\widehat{\varepsilon}_j):=
\frac{(\varepsilon-\widehat{\varepsilon}_j)^2(\Delta_j\log p)^2}{8\varepsilon}-\delta.\]
These ``raw" moments are then aggregated by a smooth function
\[\widehat{m}_{OF}^*(\varepsilon,\delta;\lambda) = \ssum_{j=1}^J\frac{\exp(\lambda\widehat{g}_{OF,j}(\varepsilon,\delta))}{\ssum_{l=1}^J
\exp(\lambda\widehat{g}_{OF,l}(\varepsilon,\delta))}\widehat{g}_{OF,j}(\varepsilon,\delta),\]
where $\lambda\in\mathbb{R}$ is a fixed, positive scalar. Note that as discussed in Section 2, this transformation approximates the maximum of $\widehat{g}_{OF,1}(\varepsilon,\delta),\dots,\widehat{g}_{OF,J}(\varepsilon,\delta)$ as $\lambda\rightarrow\infty$, but satisfies the smoothness conditions for our procedure for any finite value of $\lambda>0$.

We use a parametric bootstrap to obtain the critical value $k(1-\alpha)$, where we approximate the sampling distribution of the estimators for the respective elasticities by a joint normal distribution centered around the estimates reported in Panel A of table 1 with standard deviations equal to the respective standard errors and zero covariances. This approach can be justified by an assumption that the studies were based on mutually independent random samples from possibly different populations.

\begin{figure}[p]
\centering
\includegraphics[width=100mm, height=100mm]{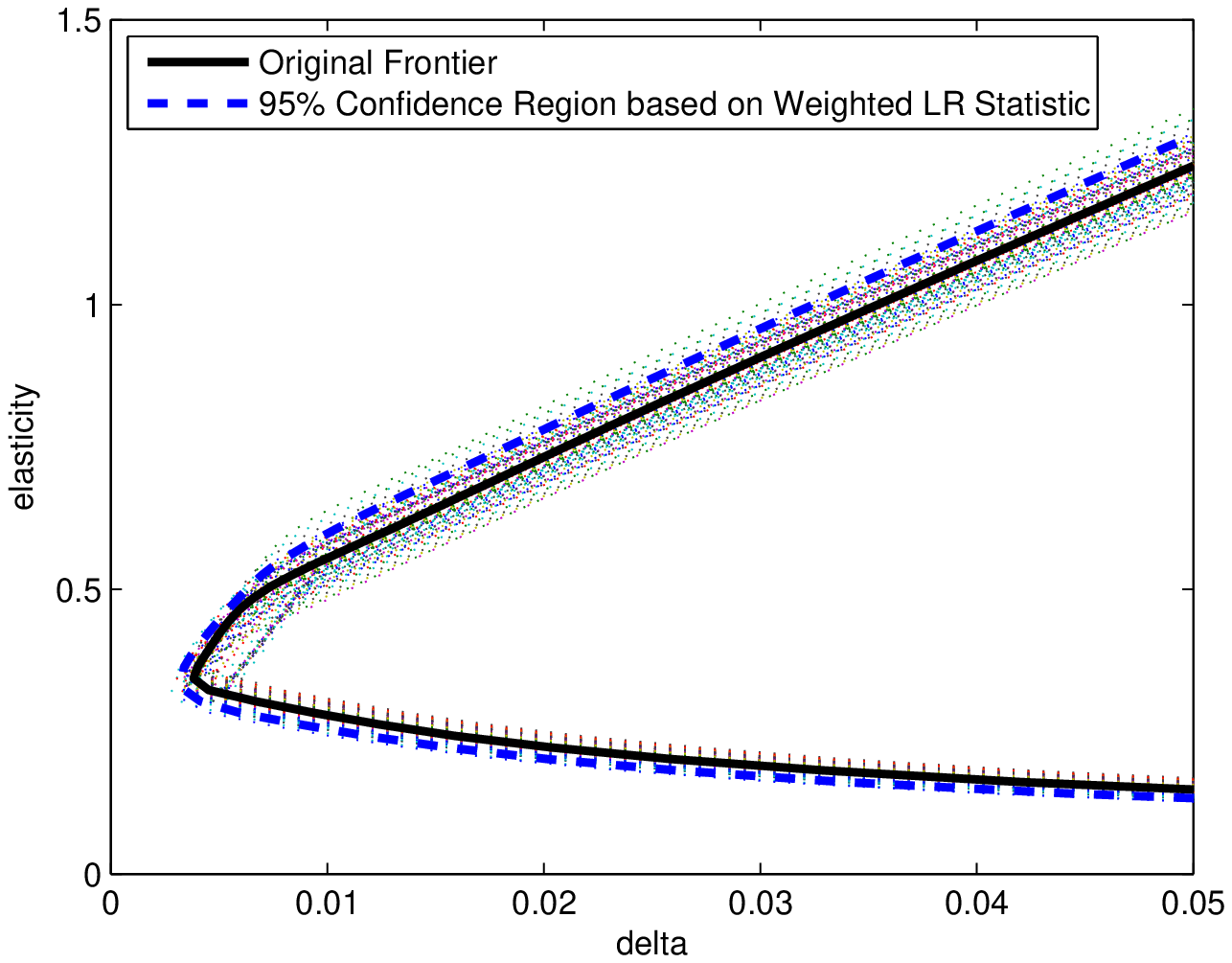}
\caption{Smoothed Bound with 95\% Confidence Set for the plausible pairs of the structural elasticity $\varepsilon$ and friction $\delta$ using the LR Statistic}\label{elast_vertical}
\includegraphics[width=100mm, height=100mm]{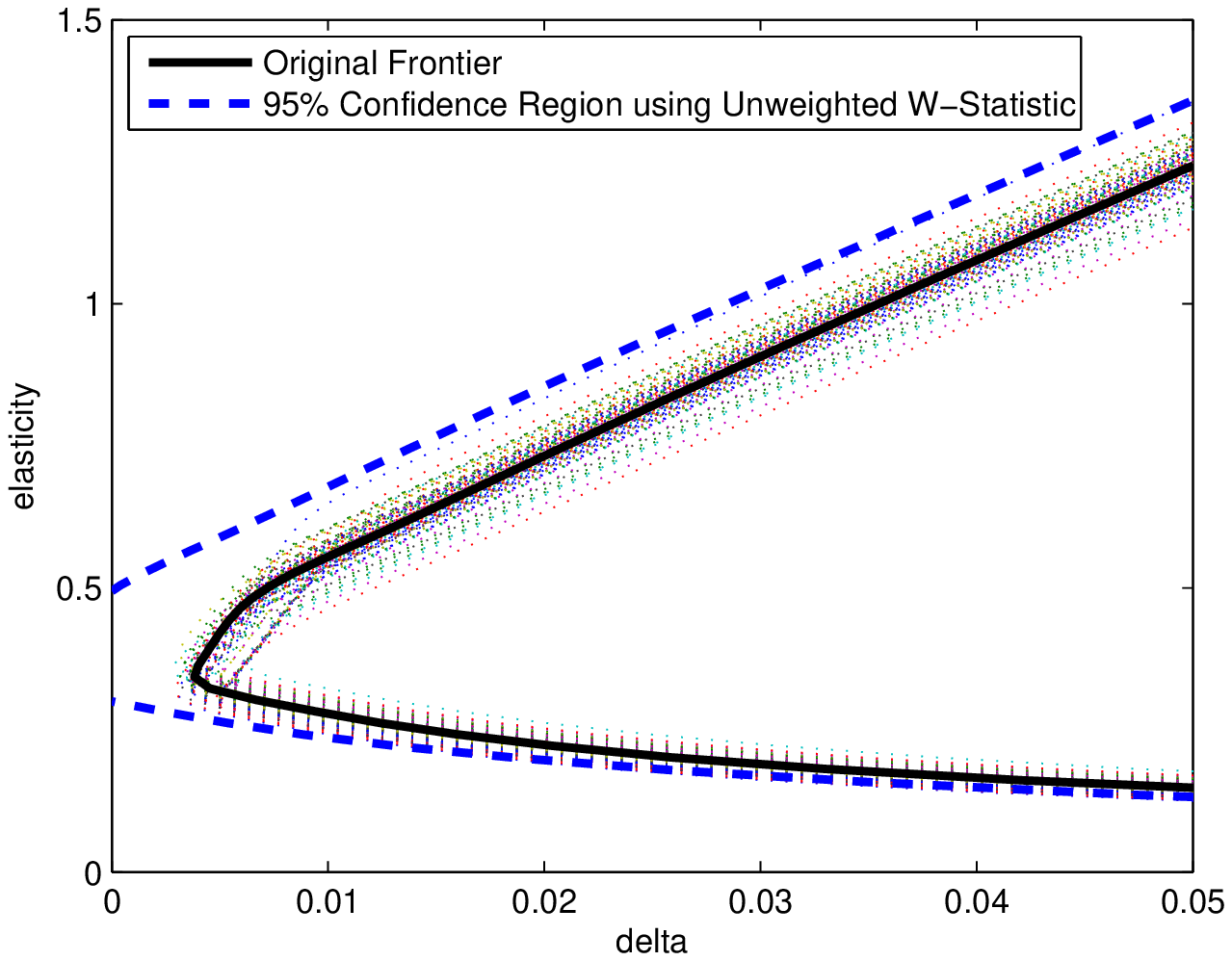}
\caption{Smoothed Bound with 95\% Confidence Set for the plausible pairs of the structural elasticity $\varepsilon$ and friction $\delta$ using Unweighted W-Statistic}\label{elast_hausdorff}
\end{figure}

Figure~\ref{elast_bound} shows that the estimated bounds coincide with the set reported in Figure 8 of \cite{C2012} except for the use of the smoothed maximum function instead of the intersection of $(\varepsilon,\delta)$-sets which leads to a slightly wider set. The 95\% confidence set based on the LR statistic\footnote{
Note that for the LR statistic we used the standard error of the negative part of $\widehat{m}^*(\theta)$ as a weighting function which improves the local approximation due to the asymmetry of the distribution for small values of $\varrho$. Note that for the $N(0,s(\theta)^2)$ distribution, the standard deviation of the negative part is proportional to $s(\theta)$, so that this weighting scheme is asymptotically equivalent to weighting by the (inverse of the) local standard deviation of $\widehat{m}^*(\theta)$.} reported in Figure~\ref{elast_vertical} is fairly narrow around the estimated bound, and does not appear to differ very much from the the collections of confidence intervals in \cite{C2012}.  Chetty presents confidence intervals that are pointwise with respect to  $\delta$, that is for each fixed value of friction $\delta$, the interval covers structural elasticity $\varepsilon$ with a prescribed probability.  In contrast, our set estimator covers all plausible values of $(\delta, \varepsilon)$ with a prescribed probability.  Thus it simultaneously performs inference on both structural elasticity $\varepsilon$ and the friction amount $\delta$. The LR confidence region is a valid joint confidence set for $(\varepsilon,\delta)$ and not only point-wise in $\delta$. Furthermore, it is not conservative in that we assume a joint sampling distribution for the elasticity estimates instead of constructing Fr\'echet-Hoeffding bounds. The LR confidence region excludes all points with $\delta\leq 0.3\%$, so that an optimization friction of at least that size would be needed to reconcile the different elasticities found in the studies considered in this meta-analysis.

Finally, we also report a 95\% confidence region based on the Wald statistic \textit{without} optimal re-weighting.\footnote{In order to adjust for the differences in order of magnitude we constructed the Hausdorff-distance based on the norm $\|(\varepsilon,\delta)\|=\sqrt{\varepsilon^2 + 100\cdot\delta^2}$. Note that in the graph the confidence region looks poorly centered around the estimated bound, but this optical impression is in fact due to the different scaling of the two axis and the difference in the slope of the frontier above and below its apex.} As in the case of HJ bounds, the shape of the resulting confidence set does not reflect the sampling variation in the estimated bounds, and the critical value for the Wald statistic is determined by perturbations of the frontier at very low values for $\varepsilon$. More importantly, in contrast to the LR-based region, the confidence set based on the Wald statistic includes points with $\delta=0$, failing to reject that the empirical elasticities can be reconciled in a  model with no optimization frictions and changing one of the main conclusions of the analysis.  Using weighted W statistics instead fixes this problem and gives a confidence set that is very similar to the LR-based confidence set; we do not report this confidence set for brevity.

\section{Conclusion}

In this paper we provide new methods for inference on parameter sets and frontiers that can be characterized by a  smooth nonlinear inequality. The proposed procedures are straightforward to implement computationally and have favorable statistical properties. By analyzing the geometric and statistical properties of different statistics, we illustrate the importance of equivariance and similarity considerations for achieving tight confidence regions. In particular, while local weighting is irrelevant for the statistical properties of the estimated frontier, it matters greatly for the size and shape of confidence sets. We also consider smoothed intersection bounds from multiple inequality restrictions, where we give an exact upper bound for the approximation error that depends only on the smoothing parameter.

We illustrate the practical usefulness of these procedures in financial econometrics with various classical examples from mean-variance analysis, including inference on Hansen-Jagannathan mean-variance sets of admissible stochastic discount factors, Markowitz-Fama mean-variance sets of admissible portfolios, and factor-based asset pricing. As a second application, we consider \cite{C2012}'s joint bounds for the elasticity of labor supply and an optimization friction. This example suggests a broader range of uses for set inference in the context of possibly misspecified or incomplete economic models.

In both examples, using invariant or precision-weighted statistics is important for maintaining major empirical conclusions that have been reached informally in prior empirical work, e.g. the inability of large values of the elasticity of intertemporal substitution to generate plausible distributions of stochastic discount factors, or the need for nontrivial optimization frictions to reconcile estimated demand elasticities from different settings. Therefore, the empirical examples illustrate our formal points about the advantages of inference based on a precision weighted metric that is invariant to parameter transformations.

\newpage
\appendix

\section{Proofs}

\subsection*{Proof of Lemma \ref{smooth_appx_bd}}

W.l.o.g., let $\max_j g_j = g_1$ and rewrite
\begin{eqnarray*}\lambda\left(\max_jg_j - m(g_1,\dots,g_J;\lambda)\right) & = &  \frac{\ssum_{j=1}^J\exp\{\lambda(g_j-g_1)\}\lambda(g_1-g_j)}{\ssum_{j=1}^J\exp\{\lambda(g_j-g_1)\}}\\
& = & \frac{\ssum_{j=1}^J\exp\{-h_j\}h_j}{\ssum_{j=1}^J\exp\{-h_j\}},
 \end{eqnarray*}
where $h_j:=\lambda(g_1-g_j)\geq0$. Clearly, this expression is nonnegative, and since $h_1=0$, the denominator is bounded from below by 1. Next note that the function
\[F(h_2,\dots,h_J):=\frac{\ssum_{j=2}^J\exp\{-h_j\}h_j}{1 +\ssum_{j=2}^J\exp\{-h_j\}}\]
is strictly quasi-concave on $\mathbb{R}_+^{J-1}$, so that the usual first-order conditions for a local extremum are sufficient for a global maximum. We can now verify that the first-order conditions for maximization of $F(h_2,\dots,h_J)$ have the symmetric solution $h_2 = \dots = h_J = h^*:= 1+W^*$ where $W^*:=W\left(\frac{J-1}e\right)$. Note that by definition of the product logarithm, $W^*=\frac{J-1}e\exp\{-W^*\}=(J-1)\exp\{-1-W^*\}$, so that
\begin{eqnarray}\nonumber\max_{h_2,\dots,h_J\geq0}F(h_1,\dots,h_J)&\equiv&\frac{\ssum_{j=2}^J\exp\{-h^*\}h^*}{1 +\ssum_{j=2}^J\exp\{-h^*\}}
=\frac{(J-1)\exp\{-1-W^*\}(1+W^*)}{1+(J-1)\exp\{-1-W^*\}}\\
\nonumber&=&\frac{W^*+(J-1)\exp\{-1-W^*\}W^*}{1+(J-1)\exp\{-1-W^*\}}=W^*=W\left(\frac{J-1}e\right)
\end{eqnarray}
Since $\lambda\left(\max_jg_j - m(g_1,\dots,g_J;\lambda)\right)=F(\lambda(g_1-g_2),\dots,\lambda(g_1-g_J))$, we therefore have that
\[\sup_{g_1,\dots,g_J}\lambda\left|\max_jg_j - m(g_1,\dots,g_J;\lambda)\right|=\sup_{h_2,\dots,h_J\geq0}|F(h_2,\dots,h_J)|=W\left(\frac{J-1}e\right)\]
which establishes the conclusion. \\[8pt]

Next, we will prove four lemmas which will be used to justify the local approximation for the Wald statistic. We consider a (stochastic or deterministic) sequence of moment functions $m_n(\theta):=m(\theta) - q_n$, where $q_n=o_p(1)$, and $m(\theta)$ satisfies Conditions R.1-R.2 from the main text, and the corresponding sequence of parameter sets $\Theta_n:=\{\theta\in\Theta:m_n(\theta)\leq 0\}$.

\begin{lemma}\label{unique_proj} Suppose the parameter space $\Theta$ is compact. Suppose that the gradient $\nabla_{\theta}m_n(\theta)$ is bounded away from zero uniformly in $\theta$ and $n=1,2,\dots$, and Lipschitz-continuous in $\theta$ with Lipschitz constant $L<\infty$. Also let $\theta_n$ be any sequence such that $\theta_n$ approaches the boundary $\Theta_n$, i.e. $d(\theta_n,\partial\Theta_n)\rightarrow0$. Then there exists $\delta>0$ such that the projection of $\theta_n$ on $\Theta_n$ is unique for all such sequences whenever $d(\theta_n,\Theta_n)<\delta$.
\end{lemma}

\textsc{Proof:} Suppose the statement wasn't true. Then for some sequence $\theta_n$, we could construct a subsequence $\theta_{k(n)}$ such that there are (at least) two distinct projections of $\theta_{s(n)}$ onto $\partial\Theta_n$ for each $n$. By compactness of $\Theta$, $\theta_{k(n)}$ has a convergent sub-subsequence $\theta_{b(n)}$ with $\lim_n\theta_{b(n)}=\theta_0$, say.
Since by construction every member of $\theta_{b(n)}$ has two distinct projections onto $\partial\Theta_n$, we can inscribe a ball of radius $r_n:=d(\theta_{b(n)},\partial\Theta_n)$ centered at $\theta_{b(n)}$ into $\Theta/\Theta_n$ such that this ball has at least two distinct points $(\theta^*_{1,b(n)},\theta^*_{2,b(n)})$ in common with $\partial\Theta_n$.

By properties of the projection, the radii of these balls corresponding to the projection points, $N(\theta^*_{j,b(n)}):=r_n^{-1}(\theta_{b(n)}-\theta^*_{j,b(n)})$ for $j=1,2$, are also normal vectors to the surface $\partial\Theta_n$ at $\theta^*_{1,b(n)}$ and $\theta^*_{2,b(n)}$, respectively. Note that, since the gradient $\nabla_{\theta}m_n(\theta)$ is bounded away from zero, we can w.l.o.g. normalize the length of the normal vectors of the surface $\partial\Theta_n$ to 1.

Note that the two points $\theta^*_{1,b(n)},\theta^*_{2,b(n)}$ are equidistant to $\theta_{b(n)}$, and therefore lie on a one dimensional sphere $S^1(r_n,\theta_{b(n)})\subset\Theta$ with center $\theta_{b(n)}$ and radius $r_n$. The curve $\eta(s)$ corresponding to a sphere of radius $r$, where $s$ is the arc length, has constant curvature $\kappa(s)\equiv\frac1{r_n}$, so that the normal vectors (with length normalized to one) $N(\theta^*_{j,b(n)}):=r_n^{-1}(\theta_{b(n)}-\theta^*_{j,b(n)})$ for $j=1,2$ satisfy $$\|N(\theta^*_{2,b(n)})-N(\theta^*_{1,b(n)})\|\gtrsim r_n^{-1}\|\theta^*_{2,b(n)}-\theta^*_{1,b(n)}\|.$$ Since $r_n\rightarrow0$, there is no upper bound on $r_n^{-1}$, so that $N(\theta)$ is not Lipschitz continuous.

However, the normal vector of $\partial\Theta_0$ at $\theta$ standardized to length 1 is given by $$N(\theta):=\|\nabla_{\theta}m(\theta)\|^{-1}\nabla_{\theta}m(\theta).$$ Condition R.2 implies that $N(\theta)$ is Lipschitz continuous in $\theta$ with a constant $1/\delta<\infty$, which leads to a contradiction. Finally, note that since that Lipschitz constant $L$ was assumed to be uniform in $\gamma$, there exists $\delta:=\delta(L)>0$ such that projection is unique whenever $r_n:=d(\theta_n,\Theta_n)<\delta$ along all sequences $\theta_n$ and $m_n(\cdot)$ \qed

\begin{lemma}\label{lem_proj}
Suppose R.1 and R.2 hold, and let $\theta_n$ be a sequence such that $d(\theta_n,\Theta_0)\rightarrow0$. Define $\bar{\theta}_n\equiv\arg\min_{\widehat{m}(\theta)\leq 0}\|\theta_n-\theta\|^2$.
Then $\bar{\theta}_n$ is uniquely defined with probability approaching 1, and satisfies $d(\bar{\theta}_n,\Theta_0)\stackrel{p}{\rightarrow}0$.
\end{lemma}

\textbf{Proof of Lemma \ref{lem_proj}.} For a given value of $\delta>0$, let $\eta:=\min\{m(\theta): d(\theta,\Theta_0)\geq\delta\}$. By continuity of $m(\theta)$ and compactness of $\Theta$, $\eta>0$. By Condition R.1 $\sup_{\theta\in\Theta_0}\widehat{m}(\theta)<\eta$ w.p.a.1, and therefore $d(\bar{\theta}_n,\Theta_0)<\delta$ with probability approaching 1. Uniqueness follows from Lemma \ref{unique_proj}, noting that the conclusion holds regardless of whether the sequence $m_n(\theta)$ is deterministic or stochastic \qed

\begin{lemma}\label{lem_proj_appx} Suppose Conditions R.1 and R.2 hold, and consider a sequence of parameter values $\theta_n\in\Theta$ such that $\theta_n\rightarrow\partial\Theta_0$. For a sequence $q_n=o_p(1)$ and $\widehat{\Theta}_n:=\{\theta\in\Theta:\widehat{m}(\theta)\leq q_n\}$, we have
\[\sqrt{n}d(\theta_n,\widehat{\Theta}_{0,\delta}) = \left[\|\nabla_{\theta}m(\theta_n)\|^{-1}(\Gn(\theta_n)+\sqrt{n}(m(\theta_n)-q_n))+ o_p(1)\right]_+ \]
In particular, for $q_n\equiv0$, we have
\[\sqrt{n}d(\theta_n,\widehat{\Theta}_{0,\delta}) = \left[\|\nabla_{\theta}m(\theta_n)\|^{-1}(\Gn(\theta_n)+\sqrt{n}m(\theta_n))+ o_p(1)\right]_+ \]
\end{lemma}

\textbf{Proof of Lemma \ref{lem_proj_appx}.} Without loss of generality, we only consider the case $q_n\equiv0$. Note that for the general case, we can apply the same argument to the functions $\widehat{m}(\theta)-q_n$ and $m(\theta)-q_n$ which inherit the properties R.1-R.2.

The projection of $\theta_n$ onto the set
$\widehat{\Theta}_{0,\delta}:=\{\theta\in\Theta^{\delta}: \widehat m(\theta)\leq 0\}$ is given by
\[\tilde{\theta}_n:=\arg\min_{\theta'\in\Theta^{\delta}:\widehat m(\theta')\leq 0}
\|\theta_n- \theta'\|^2.\]

The Lagrangian for this constrained minimization problem is $\|\theta_n-\theta'\|^2 +2 \lambda\widehat m(\theta')$, where the Kuhn-Tucker multiplier $\lambda\geq0$. Since $\delta>0$ is fixed and $d(\theta_n,\Theta_0)\rightarrow0$ it follows from Lemma \ref{lem_proj} that $\bar{\theta}_n$ is contained in the interior of $\Theta^{\delta}$ with probability approaching one. Hence for $n$ large enough,
the quantity $\bar \theta_n$ can be taken to be an interior solution of the saddle-point problem
\begin{eqnarray}
\nonumber
 (\bar{\theta}_n-\theta_n)+\nabla_{\theta} \widehat m(\bar{\theta}_n)\lambda&=&0,\\
\nonumber  \widehat m(\bar\theta_n) &=&0.
\end{eqnarray}

We can now use a mean-value expansion to obtain
\begin{eqnarray}
\nonumber
(\bar{\theta}_n-\theta_n)+\nabla_{\theta}\widehat m(\bar{\theta}_n)\lambda&=&0,\\
\nonumber \widehat{m}(\theta_n) + \nabla_{\theta}
m(\check{\theta}_n)' (\bar{\theta}_n-\theta_n)&=&0,
\end{eqnarray}
for an intermediate value $\check{\theta}_n$. By Lemma \ref{lem_proj}, $\nabla_{\theta} \widehat m(\check{\theta}_n) = \nabla_{\theta} m(\theta_n) + o_p(1)$ and
$\nabla_{\theta} m(\bar{\theta}_n) = \nabla_{\theta} m(\theta_n) + o_p(1)$. 
Hence, solving for $(\bar{\theta}_n-\theta_n)$ and applying the extended continuous mapping theorem, we obtain
\begin{eqnarray*} \bar{\theta}_n-\theta_n
& = & -[\nabla_{\theta} m(\theta_n)(\nabla_{\theta} m(\theta_n)'\nabla_{\theta} m(\theta_n))^{-1} + o_p(1)] \widehat m(\theta_n)
 \end{eqnarray*}
Now, since $\sqrt{n} (\widehat m - m) = \Gn = O_p(1)$ in $\ell^\infty(\Theta^\delta) $,  we obtain
 \begin{eqnarray*}
\sqrt{n}(\bar{\theta}_n-\theta_n)
& = & - \nabla_{\theta} m(\theta_n)(\nabla_{\theta} m(\theta_n)'\nabla_{\theta} m(\theta_n))^{-1}(\Gn(\theta_n)+\sqrt{n}m(\theta_n))+o_p(1)
 \end{eqnarray*}

Thus we have that
 \begin{eqnarray}
\nonumber\sqrt{n} d(\theta_n, \widehat \Theta_n^{\delta}) & = &\sqrt{n} \| \bar \theta_n - \theta_n \| 1\{\widehat m(\theta_n)>0\}\\
& = &   \left|(\nabla_{\theta} m(\theta_n)' \nabla_{\theta} m(\theta_n))^{-1/2}(\Gn(\theta_n)+\sqrt{n}m(\theta_n)) + o_p(1)\right|  \\
  &  \text{ }   &  \times 1 \{\Gn(\theta_n) + \sqrt{n}m(\theta_n) >0 + o_p(1)  \} \\
\nonumber& = &   [\|\nabla_{\theta} m(\theta_n)\|^{-1}(\Gn(\theta_n)+\sqrt{n}m(\theta_n)) + o_p(1)]_+\\
\end{eqnarray}
which concludes the proof  \qed

%
%

\subsection{Proof of Theorem \ref{theorem1}} \textsc{Part 1.} (\textit{Limit law of $\mathcal{L}_n$.}) Let $\Gn \:= \sqrt{n}(\widehat m - m)$. We first show that
 \begin{eqnarray*}
\mathcal{L}_n = \sup_{\theta \in \Theta_0}
\left[\sqrt{n} \widehat m(\theta)/\hat{s}(\theta) \right ]^2_+ & =_{ \text{ \ }} &  \sup_{\theta \in \Theta_0}
\left[(\Gn(\theta) + \sqrt{n} m(\theta)) /\hat{s}(\theta) \right ]^2_+  \\
 & =_{(1)} &  \sup_{\theta \in \partial \Theta_0}
\left[(\Gn(\theta) + \sqrt{n} m(\theta)) /\widehat s(\theta) + o_p(1) \right ]^2_+ \\
 & =_{(2)} &  \sup_{\theta \in \partial \Theta_0}
\left[(\Gn(\theta) + \sqrt{n} m(\theta)) / s(\theta) + o_p(1) \right ]^2_+\\
&  \rightsquigarrow_{(3)} & \sup_{\theta \in \partial \Theta_0}
\left[(\Gp(\theta) + \sqrt{n} m(\theta)) / s(\theta) \right ]^2_+,
 \end{eqnarray*}
where the steps follows from the following arguments.

To show step (1), note that
$$
 \sup_{\theta \in \Theta_0}\left[(\Gn(\theta) + \sqrt{n} m(\theta)) /\widehat{s}(\theta) \right ]^2_+
 \geq  \sup_{\theta \in \partial \Theta_0}
\left[(\Gn(\theta) + \sqrt{n} m(\theta)) /\widehat s(\theta) \right ]^2_+. $$
So we need to show that with probability approaching 1,
$$
 \sup_{\theta \in \Theta_0} \left[(\Gn(\theta) + \sqrt{n} m(\theta)) /\widehat{s}(\theta) \right ]^2_+
 \leq   \sup_{\theta \in \partial \Theta_0}
\left[(\Gn(\theta) + \sqrt{n} m(\theta)) /\widehat s(\theta) + o_p(1) \right ]^2_+. \\
$$
To show this take a sequence $\theta_n \in \Theta_0$ such that
$$
\sup_{\theta \in \Theta_0} \left[(\Gn(\theta) + \sqrt{n} m(\theta)) /\widehat s(\theta) \right ]^2_+ =
\left[(\Gn(\theta_n) + \sqrt{n} m(\theta_n)) /\widehat s(\theta_n) \right ]^2_+,
$$
where the supremum is attained by compactness of $\Theta_0$ and continuity hypotheses on $\widehat m, m$, and $\widehat s$, and
uniform positivity of $\widehat s$.  Since $\sup_{\theta \in \Theta} |\Gn| = O_p(1)$,  $\theta_n$ can be selected to obey
$$
\sqrt{n} m(\theta_n) /\widehat s(\theta_n) = O_p(1).
$$

We now show that $\theta_n$ converges to $\partial\Theta_0$: fix $\kappa>0$, and let \[\zeta(\kappa):=\sup\left\{m(\theta)/s(\theta):\theta\in\Theta_0,d(\theta,\partial\Theta_0)\geq\kappa\right\}.\]
Next, note that Condition R.2 implies that $m(\theta)<0$ for each $\theta\in\textnormal{int }\Theta_0$: suppose this wasn't true, but that $m(\theta_0)=0$ for some $\theta_0\in\textnormal{int }\Theta_0$. Then, since for $\theta\in\Theta_0$ $m(\theta)\leq0$, $\theta_0$ represents a local maximum of $m(\cdot)$ in the interior of $\Theta_0$. Since $m(\theta)$ is differentiable, this implies that $\nabla_{\theta}m(\theta_0)=0$. However, Condition R.2 implies that $\|\nabla_{\theta}m(\theta)\|>0$, a contradiction. Therefore $m(\theta)<0$ for each $\theta\in\textnormal{int }\Theta_0$, so that by continuity and compactness, $\zeta(\kappa)<0$.

By conditions R.1 and R.3, we have $\sup_{\theta\in\Theta_0}\left|\widehat{m}(\theta)/\widehat{s}(\theta)-m(\theta)/s(\theta)\right|<-\zeta$ with probability approaching one for any strictly negative value of $\zeta$. Hence the supremum of $\sqrt{n}\left[\widehat{m}(\theta)/\widehat{s}(\theta)\right]_+^2$ for values $\theta\in\Theta_0$ such that $d(\theta,\partial\Theta_0)\geq\kappa$ is equal to zero with probability approaching one, so that the supremum over all of $\Theta_0$ is attained at a value of $\theta$ such that $d(\theta,\partial\Theta_0)<\kappa$. Since $\kappa$ can be chosen arbitrarily close to zero, we can construct a sequence of positive random variables $\kappa_n = o_p(1)$ such that
$$
 d(\theta_n, \partial \Theta_0) \leq \kappa_n,
$$
with probability approaching 1. We conclude that
\begin{eqnarray*}
& & \sup_{\theta \in \Theta_0} \left[(\Gn(\theta) + \sqrt{n} m(\theta)) /\widehat s(\theta) \right ]^2_+   \\
& & \leq  \sup_{\theta \in \partial \Theta_0,  \theta + \eta \in \Theta_0, \|\eta \| \leq \kappa_n } \left[(\Gn(\theta +
\eta) + \sqrt{n} m(\theta+ \eta)) /\widehat s(\theta + \eta) \right ]^2_+.
\end{eqnarray*}
Using stochastic equicontinuity of $\Gn$ implied by R.1, the last quantity is equal to
\begin{eqnarray*}
\sup_{\theta \in \partial \Theta_0,  \theta + \eta \in \Theta_0, \|\eta \| \leq \kappa_n } \left[(\Gn(\theta) + \sqrt{n} m(\theta+ \eta)) /\widehat s(\theta + \eta) + o_p(1)\right ]^2_+,
\end{eqnarray*}
for some $o_p(1)$ term. Because $ \sqrt{n} m(\theta+ \eta) \leq 0$ for $\theta+ \eta \in \Theta_0$ and $m(\theta) = 0$ for $\theta \in \partial \Theta_0$, we conclude that the last quantity is equal to
\begin{eqnarray*}
\sup_{\theta \in \partial \Theta_0 } \left[(\Gn(\theta)/\widehat s(\theta +  \eta) +o_p(1) \right ]^2_+,
\end{eqnarray*}
for the same $o_p(1)$ term. This verifies equality (1).

Equality (2) follows from  using R.3 and that the fact that $\sup_{\theta \in \Theta } |\Gn(\theta)| = O_p(1)$ implied by R.1. Equality (3) follows from the  application of R.1 and the Continuous Mapping Theorem.  \\[4pt]

\textsc{Part 2.} (\textit{Limit Law of $\mathcal{W}_n$}). Recall that we define the set estimator $\widehat{\Theta}_{0,\delta}$ as
\[\widehat{\Theta}_{0,\delta}:=\left\{\theta\in\Theta^{\delta}:\widehat{m}(\theta)\leq0\right\}.\]
In analogy to part 1, we establish the conclusion by the following steps:
 \begin{eqnarray*}
\sqrt{\mathcal{W}_n} = \sup_{\theta \in \Theta_0}
\sqrt{n}d(\theta,\widehat{\Theta}_{0,\delta})/\widehat{w}(\theta)
 & =_{(1)} &  \sup_{\theta \in \partial \Theta_0}
\left[(\Gn(\theta) + \sqrt{n} m(\theta)) /\widehat w(\theta) + o_p(1) \right ]_+ \\
 & =_{(2)} &  \sup_{\theta \in \partial \Theta_0}
\left[(\Gn(\theta) + \sqrt{n} m(\theta)) / w(\theta) + o_p(1) \right ]_+\\
&  \rightsquigarrow_{(3)} & \sup_{\theta \in \partial \Theta_0}
\left[(\Gp(\theta) + \sqrt{n} m(\theta)) / w(\theta) \right ]_+,
 \end{eqnarray*}
where steps (1)-(3) are proven as follows:

To establish step (1), we first show that
\begin{eqnarray}\label{approx 1}
\sup_{\theta \in \Theta_0} \sqrt{n} d(\theta, \widehat{\Theta}_{0,\delta})/\widehat{w}(\theta) =
\sup_{\Theta_n} \sqrt{n} d(\theta, \widehat{\Theta}_{0,\delta})/\widehat{w}(\theta)
\end{eqnarray}
holds with probability approaching 1, where
$$
\Theta_{n} = \{ \theta \in \Theta_0: d(\theta, \partial \Theta_0)/w(\theta) \leq \kappa_n\},
$$
and $\kappa_n$ is some sequence of positive random variables converging to zero in probability, $\kappa_n = o_p(1)$. Note that right hand side of (\ref{approx 1}) is less than or equal to the left hand side of (\ref{approx 1}) by construction, so we only need to show that w.p.a.1, the right hand side can not be less. To this end, fix some $\kappa>0$ and note that using the same line of reasoning as for part 1, we have \[\zeta(\kappa):=\sup\{m(\theta):\theta\in\Theta_0,d(\theta,\partial\Theta_0)\geq\kappa\}<0.\] Furthermore, for every $\zeta<0$ we have that with probability approaching 1, $\sup_{\theta\in\Theta}|\widehat{m}(\theta)-m(\theta)|<-\zeta$, and therefore each $\theta\in\Theta_0$ with $d(\theta,\partial\Theta_0)\geq\kappa$ is included in $\widehat{\Theta}_{0,\delta}$. In that event, we can only have $d(\theta,\widehat{\Theta}_{0,\delta})>0$ for values of $\theta$ within a distance $\kappa$ of the boundary $\partial\Theta_0$. Since $\kappa$ can be chosen arbitrarily small, we can choose a sequence $\kappa_n=o(1)$ such that the right-hand side in (\ref{approx 1}) holds with probability converging to one.

Next, let $\theta_n$ be a sequence such that
$$
\sup_{\theta \in \Theta_0} \sqrt{n} d(\theta, \widehat{\Theta}_{0,\delta})/\widehat{w}(\theta)= \sqrt{n} d(\theta_n, \widehat{\Theta}_{0,\delta})/\widehat{w}(\theta_n).
$$
Note that since $d(\theta,\widehat{\Theta}_{0,\delta})$ and $\widehat{w}(\theta)$ are continuous functions of $\theta$, $\widehat{w}(\theta)$ is bounded away from zero with probability approaching 1, so that since $\Theta_0$ is compact, the supremum is attained with probability approaching 1. Since $\delta>0$ is fixed and by Lemma \ref{lem_proj} with probability approaching 1, we have $d(\bar{\theta}_n,\theta_n)/\widehat{w}(\theta)<\delta$, so that $\bar{\theta}_n$ is contained in the interior of $\Theta^{\delta}$ w.p.a.1.

It follows from Lemma \ref{lem_proj_appx} that
 \begin{equation}\label{mv_exp_thm1}
\nonumber\sqrt{n} d(\theta_n, \widehat \Theta_{0,\delta}) =
\left[\|\nabla_{\theta} m(\theta_n)\|^{-1}(\Gn(\theta_n)+\sqrt{n}m(\theta_n)) + o_p(1)\right]_+\\
\end{equation}
Hence, dividing by $\widehat{w}(\theta)$ and using stochastic equicontinuity of $\Gn(\theta)$, we can follow the same line of reasoning as in part 1 to obtain that with probability approaching 1,
\[\sup_{\theta\in\Theta_0}\sqrt{n}d(\theta,\widehat{\Theta}_{0,\delta})/\widehat{w}(\theta)
\leq\sup_{\theta \in \partial \Theta_0,  \theta + \eta \in \Theta_0, \|\eta \| \leq \kappa_n }\left[\frac{\Gn(\theta)+\sqrt{n}m(\theta+\eta)}{\|\nabla_{\theta} m(\theta)\|\widehat{w}(\theta+\eta)}+o_p(1)\right]_+\]
Noting again that $\sqrt{n}m(\theta+\eta)\leq 0$ for $\theta+\eta\in\Theta_0$, we can bound the expression by
\[\sup_{\theta\in\Theta_0}\sqrt{n}d(\theta,\widehat{\Theta}_{0,\delta})/\widehat{w}(\theta)
\leq\sup_{\theta \in \partial \Theta_0,  \theta + \eta \in \Theta_0, \|\eta \| \leq \kappa_n }\left[\frac{\Gn(\theta)}{\|\nabla_{\theta} m(\theta)\|\widehat{w}(\theta+\eta)}+o_p(1)\right]_+\]
where again the $o_p(1)$ term is the same as in the first inequality.

Therefore, we can use uniform convergence of the weighting function $\widehat w$ to $w$ from Condition R.3 and the continuous mapping theorem to obtain
\begin{equation}
\mathcal{W}_n  \rightsquigarrow \sup_{\theta \in \partial \Theta_0}\left[\left(\|\nabla_\theta m(\theta)\|w(\theta)\right)^{-1} \Gp(\theta)\right]_+^2
 \end{equation}
where the intermediate steps are analogous to equalities (2) and (3) in the proof for $\mathcal{L}_n$.\\[4pt]

\textsc{Part 3.} (\textit{Continuity of the Limit Distributions}).
The continuity of the distribution function $\mathcal{L}$ on
$(0,\infty)$ follows from \cite{Davydov98} and from the
assumption that the covariance function of $\Gp$ is non-degenerate, i.e. $\inf_{\theta\in\Theta}\textnormal{Var}(\Gp(\theta))>0$.
The probability that $\mathcal{L}$ is greater than zero is equal to the
probability that $\max_j \sup_{\theta \in \Theta } \Gp_j(\theta) >0 $,
which is greater than the probability that $\Gp_{j'}(\theta') >0$ for
some fixed $j'$ and $\theta'$, but the latter is equal to 1/2.
Therefore the claim follows.  The claim of continuity of the
distribution function of $\mathcal{W}$ on $(0, \infty)$ follows
similarly. \qed

\subsection{Proof of Corollary \ref{corollary1}}  This corollary immediately follows from the assumed conditions and from the comments given in the main text preceding the statement of Corollary 1. \qed

\subsection{Proof of Theorem \ref{theorem2}}
We have that $\Pp[\Theta_0 \subseteq R_{LR} ]  =  \Pp[ \mathcal{L}_n \leq \widehat k(1-\alpha) ]$ by the construction of the confidence region.  We then have that for any $\alpha<1/2$ that $k(1-\alpha)$ is a continuity point of the distribution function of $\mathcal{L}$, so that for any sufficiently small $\epsilon$
\begin{eqnarray*}
\Pp[ \mathcal{L}_n \leq \widehat k(1-\alpha) ]
 \leq  \Pp[ \mathcal{L}_n \leq k(1-\alpha)  + \epsilon ] +o(1)
 \to   \Pp[ \mathcal{L} \leq k(1-\alpha) + \epsilon ],   \\
\Pp[ \mathcal{L}_n \leq \widehat k(1-\alpha) ]
 \geq  \Pp[ \mathcal{L}_n \leq k(1-\alpha)  - \epsilon ] - o(1)
 \to   \Pp[ \mathcal{L} \leq k(1-\alpha) - \epsilon ].
 \end{eqnarray*}
Since we can set $\epsilon$ as small as we like and $k(1-\alpha)$ is a continuity point of the distribution function of $\mathcal{L}$, we have that
\begin{eqnarray*}
\Pp[ \mathcal{L}_n \leq \widehat k(1-\alpha) ]
 \to   \Pp[ \mathcal{L} \leq k(1-\alpha) ] = (1-\alpha).
 \end{eqnarray*}
We can conclude similarly for the W-statistic $\mathcal{W}_n$. \qed.

\subsection{Proof of Theorem \ref{root_n_consistency}} We will give the proof only for confidence regions based on the LR statistic. The arguments for the Wald-type confidence sets are completely analogous. Let $\delta_n$ be a null sequence where $\delta_n>0$ for all $n$ and $\sqrt{n}\delta_n\rightarrow\infty$. In order to show convergence with respect to Hausdorff distance, we establish that with probability approaching 1, (a) $R_{LR}\subset\Theta_0^{\delta_n}$, and (b) $\Theta_0\subset R_{LR}^{\delta_n}$, where for $\delta>0$, $A^{\delta}:=\{x\in\mathbb{R}^k:d(x,A)<\delta\}$ denotes the $\delta$-expansion of a set $A$ in $\mathbb{R}^k$.


To prove statement (a), consider a sequence $\theta_n\in\Theta\backslash\Theta_0^{\delta_n}$. We have to show that $\theta_n\notin R_{LR}$ w.p.a.1 as $n$ increases: For any fixed $\bar{\delta}>0$, let $\eta:=\inf_{\theta:d(\theta,\Theta_0)\geq\bar{\delta}}m(\theta)$. Since $\Theta$ is compact and $m(\theta)$ is continuous by R.2, it follows from the definition of $\Theta_0$  that $\eta>0$. Hence, along any sequence $\theta_n$ such that $d(\theta_n,\Theta_0)\geq\bar{\delta}$ for all $n$, we have $\sqrt{n}m(\theta_n)\rightarrow\infty$. Since $\theta\in R_{LR}$ if and only if for the critical value $k$ $\mathcal{LR}_n\leq \widehat{k}$, where $\widehat{k}$ is tight for any $n$ large enough, it follows that $\theta_n\notin R_{LR}$ w.p.a.1. Hence we can restrict our attention to (sub-) sequences for which $d(\theta_n,\Theta_0)\rightarrow0$.

Now suppose that $d(\theta_n,\Theta_0)\rightarrow0$, and $\sqrt{n}d(\theta_n,\Theta_0)\rightarrow\infty$. Let $\theta_n^*$ be the projection of $\theta_n$ onto $\partial\Theta_0$, where $\theta_n^*$ need not converge to a particular point. Then using the same steps as in the proof of Lemma \ref{lem_proj_appx} it follows from a mean-value expansion of $m(\theta)$ around $\theta_n$ and continuity of the gradient $\nabla_{\theta}m(\theta)$ that $\sqrt{n}m(\theta_n)\rightarrow\infty$. Hence the LR-statistic diverges to infinity, and $\theta_n\notin R_{LR}$ w.p.a.1.

It remains to check claim (b), namely that $\Theta_0\subset R_{LR}^{\delta_n}$. To this end, consider a sequence $\theta_n\in\Theta_0$, and let $q_n = \widehat{s}(\theta)\sqrt{\widehat{k}(1-\alpha)/n}$. Also define $\bar{\theta}_n\equiv\arg\min_{\widehat{m}(\theta)\leq q_n}\|\theta_n-\theta\|^2$, so that $d(\theta_n,\widehat{\Theta}_{0,\delta})=\|\theta_n-\bar{\theta}_n\|$. Now note that Lemma \ref{lem_proj} implies that $d(\bar{\theta}_n,\Theta_0)\stackrel{p}{\rightarrow}0$, so that we can apply Lemma \ref{lem_proj_appx} to obtain
\[\sqrt{n}d(\bar{\theta}_n,\theta_n) = [\|\nabla_{\theta} m(\theta_n)\|^{-1} (\Gn(\theta_n) + \sqrt{n}(m(\theta_n) - q_n))+o_p(1)]_+\]
By assumption R.2, the norm of $\nabla_{\theta}m(\theta_n)$ is bounded away from zero, and $\Gn(\theta_n)$ is stochastically bounded by Condition R.1. Since $\widehat{s}(\theta)$ and $\widehat{k}$ are stochastically bounded by R.3, $\sqrt{n}q_n$ is also stochastically bounded.

Noting that $m(\theta_n)\leq0$, it follows that $\sqrt{n}(\bar{\theta}_n-\theta_n) = O_p(1)$. Therefore \[\sup_{\theta\in\Theta_0}\sqrt{n}d(\theta,R_{LR})=O_p(1)\] so that for any sequence $\delta_n$ such that $\sqrt{n}\delta_n\rightarrow\infty$, we have $d(\theta,R_{LR})/\delta_n\rightarrow 0$. Hence $\Theta_0\subset R_{LR}^{\delta_n}$ w.p.a.1, which concludes the proof \qed

\subsection{Proof of Corollary \ref{corollary2}} This corollary immediately follows from the assumed conditions and Corollary 1. \qed

\subsection{Proof of Theorem \ref{theorem3}}
In what follows let $A$ denote an absolute positive constant.  We have by definition of the Kantarovich-Rubenstein metric that
$$
\mathrm{E}_{\mQ_{V^*}} [\varphi(V^*)] - \mathrm{E}_{\mQ_V} [\varphi(V)] =o_p(1) \text{ uniformly in } \varphi \in \mathrm{BL}_1(C(\Theta)).
$$
This implies that
$$
\mathrm{E}_{\mQ_{V^*}} [\varphi([V^*]^2_+)] - \mathrm{E}_{\mQ_V}  [\varphi([V]^2_+)] =o_p(1) \text{ uniformly in } \varphi \in \mathrm{BL}_1(C(\Theta)),
$$
since the composition $\varphi \circ [\cdot]^2_+ \in A \cdot \mathrm{BL}_1(C(\Theta))$ for $\varphi \in \mathrm{BL}_1(C(\Theta))$.  This further implies
that
$$
\mathrm{E}_{\mQ_{V^*}} [\varphi'(\sup_{R_n}[V^*]^2_+)] - \mathrm{E}_{\mQ_V} [\varphi'(\sup_{R_n}[V]^2_+)] =o_p(1) \text{ uniformly in } \varphi' \in \mathrm{BL}_1(\Bbb{R}),
$$
since the composition $\varphi'(\sup_{R_n}[\cdot]^2_+) \in A \cdot \mathrm{BL}_1(C(\Theta))$ for $\varphi' \in \mathrm{BL}_1(\Bbb{R})$ and $R_n$ denoting any sequence of closed non-empty subsets in $\Theta$. Therefore, by the Extended Continuous Mapping Theorem,
$$
\mathrm{E}_{\mQ_{V^*}} [\varphi'(\sup_{\widehat {\partial \Theta_0}}[V^*]^2_+)] - \mathrm{E}_{\mQ_V} [\varphi'(\sup_{\widehat {\partial \Theta_0}}[V]^2_+)] =o_p(1) \text{ uniformly in } \varphi' \in \mathrm{BL}_1(\Bbb{R}).
$$
(Note that here we compute expectations over $V^*$ and $V$ taking $\widehat{\partial\Theta}_0$ as given; note that
our bootstrap method treats $\widehat{\partial\Theta}_0$ as fixed.) Also note that any sequence of sets $R_n$ converging to a set $R$, we have that
\begin{eqnarray*}
& & |\mathrm{E}_{\mQ_V}[ \varphi'(\sup_{R_n}[V]^2_+)- \varphi'(\sup_{R}[V]^2_+)  ]| \\
& & \leq \mathrm{E}_{\mQ_V}[ |\sup_{R_n}[V]^2_+- \sup_{R}[V]^2_+| \wedge 1  ] = o_p(1)  \text{ uniformly in } \varphi' \in \mathrm{BL}_1(\Bbb{R}),
\end{eqnarray*}
since $$\sup_{R_n}[V]^2_+- \sup_{R}[V]^2_+ = o_p(1)$$ by stochastic equicontinuity of the process $V$.
Since by Condition R.1-R.3 and Theorem \ref{root_n_consistency}, $\widehat {\partial \Theta_0}$ converges
to  $\partial \Theta_0$ in the Hausdorff distance, we have by the Extended Continuous Mapping Theorem:
\begin{eqnarray*}
& & |\mathrm{E}_{\mQ_V}[ \varphi'(\sup_{\widehat {\partial \Theta_0}}[V]^2_+)- \varphi'(\sup_{\partial \Theta_0}[V]^2_+)  ]| = o_p(1)  \text{ uniformly in } \varphi' \in \mathrm{BL}_1(\Bbb{R}),
\end{eqnarray*}
where $\mQ_V$ computes the expectation over $V$, treating $\widehat {\partial \Theta_0}$ as fixed.

Combining the steps above, we conclude by the triangle inequality that:
\begin{eqnarray*}
& & |\mathrm{E}_{\mQ_{V*}}[ \varphi'(\sup_{\widehat {\partial \Theta_0}}[V^*]^2_+)- \varphi'(\sup_{\partial \Theta_0}[V]^2_+)  ]| = o_p(1)  \text{ uniformly in } \varphi' \in \mathrm{BL}_1(\Bbb{R}),
\end{eqnarray*}
which is the same as
$$
\rho_K(\mathcal{Q}_{\mathcal{S}^*}, \mathcal{Q}_{\mathcal{S}} )= o_p(1).
$$
(Note that the bootstrap random variable $\mathcal{S}^*$ is computed having fixed $\widehat{\partial\Theta}_0$.)

It is known that the convergence $\rho_K(\mathcal{Q}_{\mathcal{S}_n}, \mathcal{Q}_{\mathcal{S}}) =o(1)$,
for any sequence of laws $\mathcal{Q}_{\mathcal{S}_n}$ of a sequence of random variables $\mathcal{S}_n$ defined on probability space $(\Omega', \mathcal{F}', P_n)$ implies the convergence of the distribution function
$$
\mathrm{Pr}_{\mathcal{Q}_{\mathcal{S}_n}}[\mathcal{S}_n \leq s] = \mathrm{Pr}_{\mathcal{Q}_{\mathcal{S}}}[\mathcal{S}\leq s] +o(1),
$$
at each continuity point $(0, \infty)$ of the mapping $s \mapsto \mathrm{Pr}[\mathcal{S}\leq s]$ and also convergence
of quantile functions
$$
\inf\{s: \mathrm{Pr}_{\mathcal{Q}_{\mathcal{S}_n}}[\mathcal{S}_n \leq s] \geq p \} = \inf\{s:  \mathrm{Pr}_{\mathcal{Q}_{\mathcal{S}}}[\mathcal{S}\leq s] \geq p\} +o(1)
$$
at each continuity point $p$ of the mapping $s \mapsto \inf\{s:  \mathrm{Pr}_{\mathcal{Q}_{\mathcal{S}}}[\mathcal{S}\leq s] \geq p\}$.   Recall from Theorem 1 that the set of continuity points necessarily includes the region $(0,1/2)$.

By the Extended Continuous Mapping Theorem (see e.g. Theorem 18.11 in \cite{vaart2})
we conclude that since $\rho_K(\mathcal{Q}_{\mathcal{S}^*}, \mathcal{Q}_{\mathcal{S}}) =o_p(1)$,  we obtain the convergence in probability of the distribution function
$$
\mathrm{Pr}_{\mathcal{Q}_{\mathcal{S}^*}}[\mathcal{S}^* \leq s] = \mathrm{Pr}_{\mathcal{Q}_{\mathcal{S}}}[\mathcal{S}\leq s] +o_p(1),
$$
at each continuity point $(0, \infty)$ of the mapping $s \mapsto \mathrm{Pr}[\mathcal{S}\leq s]$ and also convergence in probability of the quantile functions
$$
\inf\{s: \mathrm{Pr}_{\mathcal{Q}_{\mathcal{S}^*}}[\mathcal{S}^* \leq s] \geq p \} = \inf\{s:  \mathrm{Pr}_{\mathcal{Q}_{\mathcal{S}}}[\mathcal{S}\leq s] \geq p \} +o_p(1),
$$
at each continuity point $p$ of the mapping $s \mapsto \inf\{s:  \mathrm{Pr}_{\mathcal{Q}_{\mathcal{S}}}[\mathcal{S}\leq s] \geq p\}$. \qed

\subsection{Proof of Proposition \ref{hj_m_c1_2_prp}} First note that $\gamma_{HJ}$ and $\gamma_{MF}$ are continuous, differentiable functions of the elements in $v,\Sigma$, and $B$, so that C.1 follows from a CLT for $\widehat{v}, \text{vec}(\widehat{\Sigma}), \text{vec}(\widehat{B})$ and the delta-rule.

Next, we check Condition C.2 for the HJ bounds. Recall that the moment function defining the HJ bound was given by
\[m_{HJ}((\mu,\sigma)',\gamma_{HJ}) = \sqrt{S_{vv}\mu^2 - 2S_{v1}\mu + S_{11}}-\sigma \]
with $S_{vv},S_{v1},S_{11}$ as defined in section 1.

Since the derivative of $m_{HJ}((\mu,\sigma)',\gamma_{HJ})$ with respect to $\sigma$ is equal to minus one, the lower bound on the norm of the gradient holds for all $\theta\in\Theta$. Also, since the eigenvalues of $\Sigma$ are bounded away from zero, and the elements of $|v|$ are bounded, $|S_{vv}|$ and $|S_{v1}|$ are also bounded. Furthermore, we can easily verify that $$S_{vv}\mu^2-2S_{v1}\mu+S_{11}\geq\frac{S_{vv}S_{11}-S_{v1}^2}{S_{vv}}.$$ for all values of $\mu$. Now, since the eigenvalues of $\Sigma^{-1}$ are bounded away from zero, and $v'v 1_N'1_N - (1_N'v)^2$ is bounded from below by a positive constant by assumption, we can also bound $S_{vv}\mu^2-2S_{v1}\mu+S_{11}$ away from zero. 

Noting that the parameter space $\Theta^{\delta}\times\Gamma$ is compact, it follows that the second derivatives of $m^{(HJ)}(\theta,\gamma_{HJ})$ with respect to $\gamma_{HJ}$ and $\theta$, respectively, are bounded. In particular, this implies Lipschitz continuity of the gradients $\nabla_{\gamma}m^{(HJ)}(\theta,\gamma_{HJ})$ and $\nabla_{\theta}m^{(HJ)}(\theta,\gamma_{HJ})$.

The arguments for the moment function $m_{MF}(\theta,\gamma_{MF})$ for the MF mean-variance set for asset portfolios are completely analogous. \qed

\subsection{Proof of Corollary \ref{corollary3}} First note that by Proposition \ref{hj_m_c1_2_prp}, Conditions C.1 and C.2 hold.
Hence, in order to prove this corollary it suffices to show that
$$
\rho_K(\mathcal{Q}_{\widehat t'Z^*}, \mathcal{Q}_{t'Z}; C(\Theta)) = o_p(1).
$$
Without loss of generality we can take $\sup_{\theta} \|\widehat t(\theta) \| \leq 1$ and $\sup_{\theta} \|t(\theta)\| \leq 1$.  The claim will follow
from
$$
\rho_K(\mathcal{Q}_{\widehat t'Z^*}, \mathcal{Q}_{t'Z}; C(\Theta)) \leq \rho_K(\mathcal{Q}_{\widehat t'Z^*}, \mathcal{Q}_{\widehat t'Z}; C(\Theta)) + \rho_K(\mathcal{Q}_{\widehat t'Z}, \mathcal{Q}_{t'Z}; C(\Theta)) = o_p(1).
$$
That $\rho_K(\mathcal{Q}_{\widehat t'Z^*}, \mathcal{Q}_{\widehat t'Z}; C(\Theta))=o_p(1)$ follows immediately
from $\rho_K(\mathcal{Q}_{Z^*}, \mathcal{Q}_{Z}) = o_p(1)$ and
$\varphi (\widehat t' \cdot) \in \mathrm{BL}_1(\Bbb{R}^k)$. Indeed,
\begin{eqnarray*}
|\varphi(\widehat t(\theta)'
z(\theta))- \varphi(\widehat t(\theta)'\bar z(\theta))|  & \leq & \sup_\theta |\widehat t(\theta)'(z(\theta)-\bar z(\theta))|\wedge 2 \\
 & \leq &  [(\sup_{\theta} \|\widehat t(\theta)\| \sup_\theta \|z(\theta) - \bar z(\theta)\|) \wedge 2]\\
  & \leq & [\sup_\theta \|z(\theta) - \bar z(\theta)\| \wedge 2].
\end{eqnarray*}
 That $\rho_K(\mathcal{Q}_{\widehat t'Z}, \mathcal{Q}_{t'Z}; C(\Theta)) = o_p(1)$ follows because uniformly in $\varphi \in \mathrm{BL}_1(C(\Theta))$
\begin{eqnarray*}
 | \Ep_{\mathcal{Q}_Z}[ \varphi (\widehat t' Z)] - \varphi (t'Z) | & \leq &  \Ep_{\mathcal{Q}_Z}[ \sup_\theta |(\widehat t(\theta) - t(\theta))'Z(\theta) | \wedge 2] \\
& \leq &   \Ep_{\mathcal{Q}_Z}[ \sup_{\theta} \| \widehat t(\theta) - t(\theta) \| \sup_{\theta }\|Z(\theta)\| \wedge 2] = o_p(1),
\end{eqnarray*}
where $\Ep_{\mathcal{Q}_Z}$ computes the expectation over $Z$, treating $\widehat t$ as fixed. \qed

\subsection{Proof of Proposition \ref{invariance_prop}.} For the LR-type statistic, it is sufficient to notice that $\mathcal{L}_n(\theta;\eta)=\mathcal{L}_n(\eta^{-1}(\eta(\theta)))=\frac{\sqrt{n}\widehat{m}_n(\theta)}{\widehat{s}(\theta)}$, and therefore only depends only on quantities evaluated at $\theta$. Hence, $$\mathcal{L}_n=\sup_{\theta\in\Theta_0}\mathcal{L}_n(\theta)=\sup_{\eta\in\eta(\Theta_0)}\mathcal{L}_n(\eta^{-1}(\eta)),$$ and $\phi_{Ln}(\theta;\eta)$ is invariant with respect to parameter transformations. Asymptotic similarity follows from Condition R.1 and the continuous mapping theorem.

Next, define $\widehat{H}_{0,\delta}=\{\eta\in\eta(\Theta^{\delta}):\widehat{m}(\eta)\leq0\}$, noting that $\widehat{H}_{0,\delta}=\eta(\widehat{\Theta}_{0,\delta})$. We now distinguish three cases regarding the limit point of the sequence $\theta_n$: (1) $\theta_0\in\textnormal{ int }\Theta_0$, the interior of $\Theta_0$, (2) $\theta\in \Theta/\Theta_0$, and (3) $\theta\in\partial\Theta_0$.

Since the weighting functions $\widehat{w}(\theta;\eta)$ were assumed to satisfy Condition R.3, Theorem \ref{root_n_consistency} implies that in case (1), $\phi_{Wn}(\theta;\eta)\rightarrow 0$ for all values of $\eta$, including the identity transformation, so that by continuity of $\mathcal{W}_n(\theta;\eta)$ in $\theta$, we have
\begin{eqnarray*}
 && \lim_{n \to \infty} \mathrm{P}(\phi_{Wn}(\theta_n;\eta)\neq\phi_{Wn}(\theta_n;\eta)) \\ && \leq   \max\{\lim_{n \to \infty}P(\phi_{Wn}(\theta_0;\eta)=1),\lim_{n \to \infty}P(\phi_{Wn}(\theta_0)=1)\} =  0.
\end{eqnarray*}
Similarly, in case (2), we have
\begin{eqnarray*}
 &&  \lim_{n \to \infty} \mathrm{P}(\phi_{Wn}(\theta_n;\eta)\neq\phi_{Wn}(\theta_n;\eta))\\ && \leq \max\{\lim_{n \to \infty}P(\phi_{Wn}(\theta_0;\eta)=0),\lim_{n \to \infty}P(\phi_{Wn}(\theta_0)=0)\}=0.
 \end{eqnarray*}
Finally, consider the third case in which $\theta_0\in\partial{\Theta}_0$: Using the expansion in Lemma \ref{lem_proj_appx}, we obtain that \begin{eqnarray}\nonumber \mathcal{W}_n(\theta_n;\eta)& = & \left[\frac{d(\eta(\theta_n),\widehat{H}_{0,\delta})}{w(\theta_n;\eta)}\right]^2
=\left[\frac{d(\eta(\theta_n),\eta(\widehat{\Theta}_{0,\delta}))}{w(\theta_n;\eta)}\right]^2\\
\label{inv_proof_exp}&=&
\left[\frac{\Gn(\theta_n)+\sqrt{n}m(\theta_n)}{\|\nabla_{\eta}m_{\eta}(\theta_n;\eta)\|
w(\theta_n;\eta)}+o_\mathrm{P}(1)\right]_+^2.\end{eqnarray}

Now suppose we split the sequence $\theta_n$ into three (possibly trivial) subsequences $\theta_{q_n},\theta_{r_n}$, and $\theta_{s_n}$, respectively, such that $\sqrt{n}m(\theta_{q_n})=O(1)$, $\sqrt{n}m(\theta_{r_n})\rightarrow\infty$, and $\sqrt{n}m(\theta_{s_n})\rightarrow-\infty$. We can now analyze the behavior of $\phi_{Wn}(\cdot)$ separately along each of these subsequences. By Theorem \ref{root_n_consistency}, \[\lim_{n \to \infty} \mathrm{P}(\phi_{Wr_n}(\theta_{r_n};\eta)=0) = \lim_{n \to \infty} \mathrm{P}(\phi_{Ws_n}(\theta_{s_n};\eta)=1) = 0\]
for all $\eta$, so that by the same arguments as before, the test is asymptotically invariant along these subsequences.

Finally, consider the sequence $\theta_{q_n}$: if $\widehat{w}(\theta;\eta)=\frac{b(\theta)}{\|\nabla_{\eta}m(\eta(\theta))\|}+o_p(1)$, the expansion in (\ref{inv_proof_exp}) together with the continuous mapping theorem implies that along the subsequence $\theta_{q_n}$, \begin{eqnarray}\nonumber\mathcal{W}_{q_n}(\theta_{q_n};\eta)
&=&\left[\frac{\Gp_{n}(\theta_{q_n})+\sqrt{q_n}m(\theta_{q_n})}{b(\theta_{q_n})}+o_p(1+\sqrt{q_n}m(\theta_{q_n}))\right]_+^2\\
\nonumber&=&\left[\frac{\Gp_{n}(\theta_{q_n})+\sqrt{q_n}m(\theta_{q_n})}{b(\theta_{q_n})}+o_p(1)\right]_+^2.\end{eqnarray} so that the leading term of this expression is a function of $\theta$ which does not depend on $\eta$.

Now note that for $\theta\in\partial\Theta_0$ we have $m(\theta)=0$, so that by this expansion and stochastic equicontinuity of $\Gp(\theta)$ we have \[\mathcal{W}_n(\eta):=\sup_{\theta\in\partial\Theta_0}\mathcal{W}_n(\theta;\eta)\rightsquigarrow \mathcal{W}:=\sup_{\theta\in\partial\Theta_0}\left[\frac{\Gp(\theta)}{b(\theta)}\right]_+^2\] where the limit does not depend on $\eta$. It follows that $\widehat{k}_{W\eta}(1-\alpha)=\widehat{k}_{W}(1-\alpha)$ for all $\eta$, so that $\phi_{Wn}(\theta;\eta)$ is asymptotically invariant along $\theta_{q_n}$ with respect to transformations $\eta\in H^*$, which establishes the second conclusion.

To establish the last claim, consider the Wald statistic in (\ref{statw}) with weighting function $\widehat{w}(\theta)=\frac{s(\theta)}{\|\nabla_{\theta}m(\theta)\|}+o_p(1)$. By Condition R.1 and the approximation in (\ref{mv_exp_thm1}), $\mathcal{W}_n(\theta)$ converges in distribution to $\max\{0,Z\}^2$, where $Z\sim N(0,1)$, for all values of $\theta\in\partial\Theta_0$ and is therefore asymptotically pivotal on the boundary of $\Theta_0$. Hence for any fixed critical level $k$, the limit $\lim_{n \to \infty}P\left(\mathcal{W}_n(\theta)>k\right)$ is constant across all values of $\theta\in\partial\Theta_0$ for any value $k$.

\normalsize

\bibliographystyle{apalike}
\bibliography{biblio}

\begin{thebibliography}{}

\bibitem[Beresteanu and Molinari, 2008]{BM2008}
Beresteanu, A. and Molinari, F. (2008).
\newblock Asymptotic properties for a class of partially identified models.
\newblock {\em Econometrica}.

\bibitem[Bohrer, 1973]{Boh1973}
Bohrer, R. (1973).
\newblock An optimality property of {Scheff{\'e}} bounds.
\newblock {\em Annals of Statistics}.

\bibitem[Britten-Jones, 1999]{BJ1999}
Britten-Jones, M. (1999).
\newblock The sampling error in estimates of mean-variance efficient portfolio
  weights.
\newblock {\em Journal of Finance}.

\bibitem[Casella and Strawderman, 1980]{CSt1980}
Casella, G. and Strawderman, W. (1980).
\newblock Confidence bands for linear regression with restricted predictor
  variables.
\newblock {\em JASA}.

\bibitem[Chernozhukov et~al., 2007]{CHT2007}
Chernozhukov, V., Hong, H., and Tamer, E. (2007).
\newblock Estimation and confidence regions for parameter sets in econometric
  models.
\newblock {\em Econometrica}.

\bibitem[Chetty, 2012]{C2012}
Chetty, R. (2012).
\newblock Bounds on elasticities with optimization frictions: A synthesis of
  micro and macro evidence on labor supply.
\newblock {\em Econometrica}.

\bibitem[Cochrane, 2005]{cochrane}
Cochrane, J.~H. (2005).
\newblock {\em Asset Pricing}.
\newblock Princeton University Press.

\bibitem[Critchley et~al., 1996]{CMS1996}
Critchley, F., Marriott, P., and Salmon, M. (1996).
\newblock On the differential geometry of the wald test with nonlinear
  restrictions.
\newblock {\em Econometrica}.

\bibitem[Davydov et~al., 1998]{Davydov98}
Davydov, Y., Lifshits, M., and Smorodina, N. (1998).
\newblock {\em Local Properties of Distributions of Stochastic Functionals}.
\newblock American Mathematical Society, Providence, RI.

\bibitem[Fama, 1996]{F1996}
Fama, E. (1996).
\newblock Multifactor portfolio efficiency and multifactor asset pricing.
\newblock {\em The Journal of Financial and Quantitative Analysis}.

\bibitem[Gibbons et~al., 1989]{GRS1989}
Gibbons, M., Ross, S., and Shanken, J. (1989).
\newblock A test on the efficiency of a given portfolio.
\newblock {\em Econometrica}.

\bibitem[Hansen and Jagannathan, 1991]{HJ1991}
Hansen, L.~P. and Jagannathan, R. (1991).
\newblock Implications of security market data for models of dynamic economies.
\newblock {\em The Journal of Political Economy}.

\bibitem[Hansen and Singleton, 1982]{HS1982}
Hansen, L.~P. and Singleton, K. (1982).
\newblock Generalized instrumental variables estimation of nonlinear rational
  expectations models.
\newblock {\em Econometrica}.

\bibitem[Kaido and Santos, 2011]{KS2011}
Kaido, H. and Santos, A. (2011).
\newblock Asymptotically efficient estimation of models defined by convex
  moment inequalities.
\newblock working paper, BU and UCSD.

\bibitem[Lehmann and Romano, 2005]{LRo2005}
Lehmann, E. and Romano, J. (2005).
\newblock {\em Testing Statistical Hypotheses}.
\newblock Springer.

\bibitem[Lettau and Ludvigson, 2009]{LL2009}
Lettau, M. and Ludvigson, S. (2009).
\newblock Euler equation errors.
\newblock {\em Review of Economic Dynamics}.

\bibitem[Ludvigson, 2012]{Lud2012}
Ludvigson, S. (2012).
\newblock Advances in consumption-based asset pricing: Empirical tests.
\newblock {\em in: George M. Constantinides and Milton Harris and Rene M. Stulz
  (eds.), Handbook of the Economics of Finance, vol. 2}.

\bibitem[Markowitz, 1952]{M1952}
Markowitz, H. (1952).
\newblock Portfolio selection.
\newblock {\em The Journal of Finance}.

\bibitem[Molchanov, 1998]{M1997}
Molchanov, I.~S. (1998).
\newblock A limit theorem for solutions of inequalities.
\newblock {\em Scandinavian Journal of Statistics}.

\bibitem[Naiman, 1984]{Nai1984}
Naiman, D. (1984).
\newblock Optimal simultaneous confidence bounds.
\newblock {\em Annals of Statistics}.

\bibitem[{Pe\~{n}aranda} and Sentana, 2010]{PS2010}
{Pe\~{n}aranda}, F. and Sentana, E. (2010).
\newblock Spanning tests in return and stochastic discount factor mean-variance
  frontiers: A unifying approach.
\newblock working paper, UPF and CEMFI.

\bibitem[Politis and Romano, 1994]{politis}
Politis, D.~N. and Romano, J.~P. (1994).
\newblock Large sample confidence regions based on subsamples under minimal
  assumptions.
\newblock {\em The Annals of Statistics}.

\bibitem[{Scheff\'{e}}, 1953]{Sch1953}
{Scheff\'{e}}, H. (1953).
\newblock A method for judging all contrasts in the analysis of variance.
\newblock {\em Biometrika}.

\bibitem[Sentana, 2009]{S2009}
Sentana, E. (2009).
\newblock The econometrics of mean-variance efficiency tests: a survey.
\newblock {\em Econometrics Journal}.

\bibitem[van~der Vaart, 1998]{vaart2}
van~der Vaart, A.~W. (1998).
\newblock {\em Asymptotic Statistics}.
\newblock Cambridge University Press.

\bibitem[van~der Vaart and Wellner, 1996]{vaart}
van~der Vaart, A.~W. and Wellner, J.~A. (1996).
\newblock {\em Weak Convergence and Empirical Processes}.
\newblock Springer-Verlag New York.

\end{thebibliography}

\newpage


 \newpage


\newpage

\end{document}